\documentclass[preprint,showpacs,preprintnumbers,amsmath,amssymb]{revtex4-1}
\usepackage{graphicx}
\usepackage{bm}
\usepackage{color}
\usepackage{amssymb}
\usepackage{epsfig}
\begin{document}

\title{Spectral statistics of multi-parametric Gaussian ensembles with chiral symmetry}
\author{Triparna Mondal and Pragya Shukla}
\affiliation{ Department of Physics, Indian Institute of Technology, Kharagpur-721302, West Bengal, India }
\date{\today}

\widetext

\begin{abstract}

  The statistics of chiral matrix ensembles with uncorrelated but multivariate Gaussian distributed elements is intuitively expected to be driven by many parameters. Contrary to intuition, however, our theoretical analysis
 reveals the existence of a  single parameter, a function of all ensemble parameters, which governs the  dynamics of spectral statistics. 
The analysis not only  extends the  similar formulation (known as complexity parameter formulation) for  the Hermitian ensembles without chirality  to those with it  but also  reveals the underlying connection between chiral complex systems  with seemingly different system  conditions as well as to other complex systems e.g.  multi-parametric  Wishart ensembles as well as generalized Calogero Sutherland Hamiltonian (CSH).
  
\end{abstract}

\maketitle
.

\section{Introduction}

For systems in which the relevant behavior is governed by a linear operator,  it is useful to consider the matrix-representation in a symmetry preserving basis which in turn puts the constraints on the type of matrix elements and/or structure of the matrix \cite{dy, zirn, psijmp} (referred as matrix constraints).  The underlying complexity in real systems however often manifests through fluctuations of physical properties making it necessary to consider their statistical behavior \cite{me,fh,gmw}. This in turn requires an analysis of  not only a single matrix but rather their  ensemble, with latter's choice sensitive to the system  specific conditions e.g.  hopping range, dimensionality, boundary conditions etc; the conditions on the choice of ensemble i.e its parameters as well as nature of randomness is referred as ensemble constraints.  The latter can conspire with matrix constraints in multiple ways to give rise to different types of statistical behavior. This motivates the present study in which the primary focus  is to analyze the influence  of a specific {\it matrix constraint}, namely chiral symmetry, on the statistical behavior of complex systems with varying {\it ensemble constraints} e.g disorder \cite{psijmp}. The motivation comes not only from the fundamental aspect of the topic but also from a range of applications in which complexity appears hand in hand with chirality e.g. charge transport in graphene \cite{fm}, spectral fluctuations in QCD Dirac operators \cite{vb,gade}, conductance fluctuations in mesoscopic systems \cite{sn,zirn}, topological systems etc \cite{been-m, chal, smb, ek}.

Consider a complex system with chiral symmetry described by an ensemble of chiral Hermitian matrices. For special cases in which the complexity subjects the matrix elements to independent and identical distributions (e.g. cases with ergodic dynamics in the basis space) thus resulting in minimum number of ensemble constraints,  the system can then be represented by a basis-invariant chiral ensemble e.g chiral Gaussian ensemble invariant under orthogonal, unitary or symplectic transformation   (referred as Ch-GOE, Ch-GUE and Ch-GSE respectively) \cite{fh,gmw,me,vb}.   For generic cases however  the information about inhomogeneity of system conditions appears through ensemble parameters (e.g. those with localized dynamics in basis space), and as a consequence an appropriate ensemble representation depends on many of them. Any variation of the system conditions changes  the ensemble parameters, thus  leading to a multi-parametric evolution of the matrix ensemble  (in a fixed basis) and it is natural to wonder whether any universality classes can be identified during non-equilibrium stages.  As revealed by  previous studies, the answer is in affirmative at least in the case of the multi-parametric, non-chiral, Hermitian ensembles  (of real-symmetric, complex Hermitian or real quaternion matrices); the reasoning is based on a common mathematical formulation of their statistical properties in which ensemble details enter only through a single function of all distribution parameters. Using the function, referred as the complexity parameter,  the non-chiral  ensembles can be mapped to  a single parametric Brownian ensemble of corresponding symmetry-class (real, complex Hermitian or real-quaternion) \cite{psalt, psand, psco, pssymp, psvalla, pswf}. The latter can be described as an ensemble of  Hermitian matrices $H=H_0 + \sqrt{Y} \; V$, with $H_0$  taken from one of the stationary ensemble of Hermitian matrices, subjected  to  perturbation $V$ taken from another stationary ensemble  and $Y$ as the perturbation parameter \cite{psbe, fkpt, sp}. The mapping not only reveals the underlying universality among non-equilibrium (non-stationary or basis-dependent) Hermitian ensembles but also helps the application of all available information for the latter to  former.  A similar formulation in case of chiral Hermitian matrices  is very desirable as well as intuitively expected but is not technically obvious; this is because their off-diagonal blocks are in general non-Hermitian.  This motivates us to analyze the  multi-parametric Gaussian ensembles of chiral Hermitian matrices with and without time-reversal symmetry and seek a single parametric formulation of their spectral and strength (i.e eigenfunctions) fluctuations. As discussed later,  the diffusion equation for the ensemble density (joint probability density function of the matrix elements) in  terms of the complexity parameter turns out to be analogous to that of a chiral Brownian ensemble (Ch-BE), a perturbed stationary chiral ensemble with its diffusion governed by the perturbation parameter\cite{chbe}; a direct diagonalization of the diffusion equation then leads to analogous diffusion equations for the spectral as well as strength statistical measures. Some of the fluctuations measures for the  Ch-BE are theoretically analyzed in \cite{chbe}, with their formulation expressed in terms of perturbation parameter. By replacing the latter by the complexity parameter, the information can then directly be used for the multi-parametric Gaussian ensembles of chiral Hermitian matrices.

The implications of the single parametric formulation of the statistics for  the multi-parametric chiral ensembles are many e.g. it reveals (i)  analogy among the statistics of different complex systems, represented by the ensembles of different ensemble constraints but same matrix constraints, 
(ii)  analogy of the statistics of a  complex system for different system conditions, (iii) the connection to a variant of Calogero-Sutherland Hamiltonian (CSH) thus providing further evidence supporting the claim  that the CSH is  hidden backbone Hamiltonian of the world of  complex systems \cite{psuni}, 
(iii) the possibility of a similar formulation for multi-parametric Wishart ensembles.
 The importance of these connections as well as the implications  makes it necessary to verify our theoretical predictions and is primary focus of the present work. For this purpose, we numerically analyze the spectral statistics of the four Gaussian chiral ensembles with different functional dependence of the distribution parameters.

The paper is organized as follows. The section II describes the diffusion of the multi-parametric probability density of the chiral ensemble under consideration and  presents the complexity parametric formulation of its diffusion when  the ensemble parameters (a few or all) are varied.   As the steps are essentially the same as in non-chiral case discussed in \cite{psand}, we avoid repetition here and only mention the diffusion equation for the ensemble density. An exact diagonalization of the latter then leads to complexity parameter driven diffusion equations for the joint probability distribution functions of the eigenvalues and eigenfunctions.  The relevant steps are described in section III; here again we mention only those steps which are different from the non-chiral cases. The numerical analysis presented in section IV verifies our theoretical predictions.  We conclude in section V with a brief summary of our results and open questions.

\section{Multi-parametric Gaussian ensembles with chirality and Hermitian constraints}

\subsection{Matrix-representation}

A $(2N+\nu) \times (2N+\nu)$ Hermitian matrix with chirality constraint can be described as 
\begin{eqnarray}
H= \left( {\begin{array}{cc}   0  & C  \\  C^{\dagger} & 0 \end{array}} \right) .
\label{ch1}
\end{eqnarray}
where $C$ is a general a $N\times(N+\nu)$ complex matrix if $H$ has no other anti-unitary symmetry; as clear from above, $H_{k,N+l} = C_{kl}$ \cite{zirn, fh, me, psijmp}. For cases with time-reversal  symmetry also present, $C$ is a real or quaternion matrix based on the presence/ absence of   rotational symmetry (i.e integer or half integer angular momentum). For clarity purposes, here we confine our study only to $C$ real or complex with no other matrix constraints. The elements of  $C$ matrix  can then be written as $C_{kl}=\sum_{s=1}^{\beta} (i)^{s-1}C_{kl;s}$ where $k=1 \to N, l=1 \to (N+\nu)$ and $\beta=1$ or $2$ for $C$ real or complex. (The generalization to quaternion $C$ can be done following similar steps but is technically tedious and is therefore not included here).

\subsection{Diffusion of matrix elements: ensemble complexity parameter}

Using eq.(\ref{ch1}), the distribution, say $\rho(H)$, of the elements of the matrix $H$ can be expressed in terms of those of $C$:
\begin{eqnarray}
\rho(H) = \rho_c(C) \; F_c \; F_h
\label{rhoh}
\end{eqnarray}
with $\rho_c(C)$ as the probability density of the ensemble of $C$ matrices, with $F_c$ and $F_h$ as the constraints due to  chirality and Hermiticity of $H$, respectively:  

$F_c=\left(\prod_{k,l=1}^N \; \delta(H_{kl}) \right)\; \left(\prod_{k=1}^{N} \prod_{l=1}^{N+\nu}\delta(H_{k,N+l}-C_{k,l})\right)$ and $F_h(H)=\delta(H-H^{\dagger})$.

For simple presentation of our formulation, here we consider elements of $C$ as  independent Gaussian distributed, with  arbitrary mean and variances: 
\begin{eqnarray}
\rho_c(C; h,b)  &=&  \mathcal{N} \;  {\rm exp}\left[-  \sum_{k,l,s} \;{1\over 2 h_{kl;s}} \left( C_{kl;s} -b_{kl;s} \right)^2 \right]
\label{rhoc}
\end{eqnarray}
with $\sum_{k,l,s}  \equiv \sum_{k=1}^N \sum_{l=1}^{N+\nu} \sum_{s=1}^{\beta}$ and  $\mathcal{N} $ as a normalization constant.  Here $h \equiv \left[h_{kl,s} \right]$ and $b\equiv \left[b_{kl,s} \right]$ refer to the matrices of variances and mean values of $C_{kl;s}$. Clearly, with different choices of $h$ and $b$-matrices, eq.(\ref{rhoc}) can give rise to many chiral ensembles; some of them are used later in section IV for numerical verification of our results.

Using $H_{k,N+l} = C_{kl}$, eq.(\ref{rhoc}) leads to the ensemble density $\rho(H)$ of the $H$- matrix 
\begin{eqnarray}
\rho(H; h,b) &=&  \mathcal{N} \;  {\rm exp}\left[-   \sum_{k,l,s} \; {1\over 2 h_{kl;s} } \left( H_{k(N+l)} -b_{kl;s} \right)^2 \right]  \; F_c \; F_h 
\label{rhoc2}
\end{eqnarray}

We now consider a diffusive dynamics in the ensemble-space of $C$-matrices by  a smooth variation of the parameters $h_{kl;s}$ and $b_{kl;s}$.  As the dynamics occurs in $C$-matrix space, it preserves the chirality of $H$. 
Proceeding as discussed in \cite{psco, psvalla} for non-chiral case, it can be shown that the diffusion 
depends on the multiple parameters $h_{kl;s}$, $b_{kl;s}$  only through a function $Y$, latter referred as the ensemble complexity parameter, if the matrix-basis preserves the global constraints on the system.  The single parametric evolution of the $\rho_c(C)$ can  be described as 
\begin{eqnarray}
\frac{\partial \rho_c}{\partial Y} =  \sum_{k,l,s} \frac{\partial}{\partial C_{kl;s}}\left[ \frac{\partial  \rho_c}{\partial C_{kl;s}}+\gamma \; C_{kl;s} \; \rho_c \right]
\label{ch8}
\end{eqnarray}
with  
\begin{eqnarray}
Y=-\frac{1}{2 M\gamma} \; {\rm ln} \left[\prod_{k,l}' \prod_{s=1}^{\beta} |x_{kl;s} | \;|b_{kl;s}|^2\right]+{\rm const}
\label{y}
\end{eqnarray}
where $x_{kl;s} = 1- 2 \; \gamma  \; h_{kl;s}$ and $\prod_{k,l}' $  implies a product over non-zero $x_{kl;s},b_{kl;s}$, with M as their total number; (for example for the  case with all  $x_{kl;s} \not=0$ but $b_{kl;s}=0$, we have $M=\beta N (N+\nu)$ and for case with all  $x_{kl;s} \not=0$ and $b_{kl;s} \not=0$, we have $M=2\beta N (N+\nu)$). Further $\gamma$ is an arbitrary parameter, related to final state of the ensemble  (giving the variance of  matrix elements at the end of the evolution) and  the constant in eq.(\ref{y}) is determined by the initial state of the ensemble. 

Substitution of $C_{kl;s} =H_{k N+l;s}$ in the above then leads to the evolution equation for $\rho(H)$
\begin{eqnarray}
\frac{\partial \rho}{\partial Y} =  \sum_{k,l,s} \frac{\partial}{\partial H_{k N+l;s}}\left[ \frac{\partial \rho}{\partial H_{k N+l;s}}+\gamma \;  H_{k N+l;s} \; \rho \right]
\label{chi8}
\end{eqnarray}

Eq.(\ref{chi8}) describes the  diffusion of $\rho(H)$  with a finite drift, starting from an arbitrary initial condition, say $\rho_0(H,Y_0)$ at $Y=Y_0$ and approaching the steady limit of CH-GOE/GUE as $Y \to \infty$.
With  system information in eq.(\ref{chi8}) appearing only through $Y$, its solution $\rho(H|H_0)$  remains same for different ensembles, irrespective of the details of $h$ and $b$ matrices,  if they share same $Y$-value and are subjected to same global constraints. The latter condition can be explained as follows. A generic  transformation  maps $M$ independent variables (i.e  sets $h$ and $b$) to another set $\{Y,Y_2,..,Y_M\}$ of independent variables:
$h_{kl;s}=h_{kl;s}(Y, Y_2,..,Y_M)$ and $b_{kl;s}=b_{kl;s}(Y,Y_2,..,Y_M)$. 
This transforms $\rho(h,b) \rightarrow \rho(Y, Y_2,..Y_M)$. 
Eq.(\ref{chi8})  describes the $Y$-governed evolution of $\rho$ while $Y_j$, $j=2 \to M$ remain constant.  As  discussed in \cite{psijmp, psco}, these $M-1$ constants 
can be chosen in terms of the basis constants and initial conditions if the basis (chosen to represent $H$) is kept unchanged during the evolution.  The statistics during the transition is then governed by $Y$ only  with $Y_j$, $j=2 \to M$ appearing as the  constants of evolution. For analysis of physical properties, it is important to choose a physically motivated basis. For comparison of ensembles subjected to same global constraints, the appropriate basis is the one which preserves these constraints. This in turn ensures same evolution constants $Y_j$, $j=2 \to M$ as well as a common initial state for  the ensembles under consideration; (the related examples for non-chiral cases are discussed  in \cite{psijmp, psco, psand}) and can be generalized for chiral ones). 


In \cite{psco}, a similar diffusion equation of the matrix elements confined by harmonic potential and governed by a  single parameter was also derived   for non-chiral Hermitian ensemble of multi-parametric Gaussian ensembles  (also see \cite{psvalla} for an alternative approach).  Eq.(\ref{chi8}) is different from its non-chiral counterparts only in terms of the distinct matrix elements which appear in the equation: the former has contributions only from the elements in  one off-diagonal block  $C$ (or $C^{\dagger}$) of $H$ matrix while the later has it from all off-diagonals.

A chiral Brownian ensemble (Ch-BE) is a special case of the multi-parametric chiral Gaussian ensembles; ( its ensemble parameters are same as given in eq.(\ref{ch23}) for chiral Rosenzweig-Porter ensemble). 
The  equation governing evolution of the matrix elements of Ch-BE is derived in \cite{chbe}  (see eq.(32) therein) and is analogous, as expected, to eq.(\ref{chi8}).  Various statistical measures for the former are also discussed in \cite{chbe}, and as discussed below, can directly be applied to  the multi-parametric chiral Gaussian ensembles following the analogy.





\section{Spectral Statistics}

With $H$ given by eq.(\ref{ch1}), let $E$ be its eigenvalue matrix ($E_{mn} =e_n \delta_{mn}$) and  $U$ as the eigenvector matrix, with $U_{kn}$ as the $k^{th}$ component of the eigenvector $U_n$ corresponding to eigenvalue $e_n$. Following from eq.(\ref{ch1}), Tr($H$) is zero which then implies that the  eigenvalues  of $H$ exist  in equal and opposite pairs or are zero; let us refer such pairs as ${e}_n, e_{n+N}$ with $e_n=- e_{n+N}$, $1\le n \le N$.  Clearly, with $E$ as $(2N+\nu)\times (2N+\nu)$ diagonal matrix,  the number of zero eigenvalues is $\nu$. Henceforth, the eigenvalues are labelled such that $e_k$, $k=1 \to N$ correspond to positive eigenvalues with their negative counterparts lying from $k=N+1 \to 2N$ and $k=2N+1 \to 2N+\nu$ refers to zero eigenvalues.

A variation of system conditions perturbs $H$, resulting in dynamics of the matrix elements and thereby of the eigenvalues and eigenfunctions.  The latter's response  to change in $H$ can be derived from the eigenvalue equation $H U=U E$ along with the unitary condition $U^{\dagger}.U=I_{2N+\nu}$; this has been discussed in detail for non-chiral cases in many previous studies   \cite{psalt, pswf, pssymp, psvalla, psco} and in \cite{chbe} for chiral Brownian ensemble.  Although the intermediate steps in chiral cases are essentially similar to the non-chiral ones,  their final responses turn out to be different. The  difference mainly arises from the response of the pairwise symmetric  eigenvalues, chirality induced relations between eigenfunction components as well as existence of zero modes in a chiral matrix.
As  the present work is confined to  the spectral fluctuation analysis,   we include chiral spectral responses  in {\it appendix A} to make the presentation self content.

\subsection{Joint Probability distribution of eigenvalues}

As in the non-chiral case \cite{psalt, pswf}, an exact diagonalization of eq.(\ref{chi8}) leads to the diffusion equations for the eigenvalues and eigenfunctions.  These equations can also be obtained by standard second order perturbation theory for Hermitian matrices with chiral symmetry. With primary focus on the eigenfunction statistics, the perturbation route was used in \cite{chbe}  in case of Ch-BEs \cite{chbe}. This  route however is based on $H$ expressed as the sum of two matrices (the elements of one matrix subjected to perturbation by the other which in turn manifests as perturbation of one stationary ensemble by another) and its application to multi-parametric case (where the parameters of a single ensemble are subjected to perturbation) is not directly obvious. It is therefore instructive to consider the exact diagonalization route too; (it also gives  insights about how the dynamics of $\rho$ in the matrix space is mimicked by that in the parameter space).

Let us define $P(E) \equiv P(e_1, e_2, \ldots,  e_{2N+\nu})$ as the joint probability density function (JPDF) of the eigenvalues $e_i$, $i=1,2,\ldots, 2N+\nu$ of $H$:
\begin{eqnarray}
P(e_1, e_2, \ldots,  e_{2N+\nu})  \equiv P_N(e_1, e_2, \ldots,  e_N) \; \prod_{k=1}^N  \delta(e_k + e_{k+N}) \; \prod_{n=1}^{\nu} \delta(e_{2N+n})
\label{pe1}
\end{eqnarray}
with $P_N(E) \equiv P_N(e_1, e_2, \ldots,  e_N)$ as the JPDF of the $N$ non-zero positive  eigenvalues. As discussed in {\it appendix B}, an exact integration of eq.(\ref{chi8}) leads to $Y$ governed evolution of $P_N$. (As the  derivation is similar to that of non-chiral cases (see \cite{psco, psalt},  the intermediate steps  are  discussed in {\it appendix B} to avoid repetition).

\begin{eqnarray}
\frac{\partial P_N}{\partial Y}=2 \sum_{n=1}^N \frac{\partial}{\partial e_n}\left[\frac{\partial}{\partial e_n}-\beta \left(\frac{\nu+1/2}{ e_n}+\sum_{m=1}^N \frac{2 e_n}{e_n^2- e_m^2} \right)+ \gamma e_n\right]P_N
\label{ch19}
\end{eqnarray}
The above equation describes the diffusion of $P_N(E,Y)$, with a finite drift, from an arbitrary initial state $P_N(E_0,Y_0)$ at $Y=Y_0$ . In limit $\frac{\partial P_N}{\partial Y} \rightarrow 0$ or $Y\rightarrow \infty$, the diffusion approaches a unique steady state: $P_N(E; \infty) = C_{\beta} \; |Q_N|^{\beta}$ with $C_{\beta}$ as the normalization constant and
\begin{eqnarray}
|Q_N|^{\beta}= \prod_{m < n=1}^N |e_m^2 -e_n^2|^{\beta} \; \prod_{k=1}^N |e_k|^{ \beta (\nu+1/2)} \; \; {\rm exp}\left[-{\gamma \over 2 } \sum_{k=1}^N e_k^2\right]
\label{qn}
\end{eqnarray}

It is desirable to seek the solution of eq.(\ref{ch19}) for finite $Y$. This is however technically difficult, requires a separate study and is also beyond the purview of the present study where our main objective is to numerically confirm the validity of complexity parametric formulation of the spectral statistics. Important insights in the latter can however be obtained by following analogy: eq.(\ref{ch19}) for the multi-parametric case is analogous to  that of Ch-BE \cite{chbe}; (this is as expected, following  the analogy of eq.(\ref{chi8}) with evolution of the ensemble density of a Ch-BE). Consequently, the theoretical results, obtained in \cite{chbe}), for the statistics of chiral Brownian ensembles can  directly be used for the multi-parametric chiral Gaussian ensembles. For example,  following the same steps as discussed in \cite{chbe}, the diffusion equation for $P_N(E,Y)$ can be rewritten in terms of the Schrodinger equation for the general state $\Psi(E,Y)={P_N(E,Y)\over |Q|^{\beta/2}}$ of a variant of the Calogero-Sutherland (CS) Hamiltonian of the interacting particles (with eigenvalues now playing the role of particles). The ground state and many of the excited states of the standard CSH (and many variants) have already been worked out and relevant information about its particle correlations is available.The information can then be used in deriving the solution for $P_N(E,Y)$ and the  spectral correlations for the present case. (A similar connection between multi-parametric Gaussian ensembles without chirality and standard CSH  has  been used in past to derive the spectral correlations for the former.  Although the steps remain essentially same for the two cases  but the difference in confining potential is expected to manifest in  long range correlations). Alternatively, for the Ch-BE, a hierarchical set of equations for the spectral correlations is derived in \cite{chbe} (see eq.(77) therein) and can directly be applied for the multi-parametric chiral Gaussian ensembles (with $Y$ in eq.(76) of \cite{chbe} now replaced by $Y$ in eq.(\ref{y}).

Further insights about the spectral statistics in the case of multi-parametric chiral case can also be gained by a comparison of eq.(\ref{ch19}) with its non-chiral counterpart derived in \cite{psalt, psco, psvalla} (see eq.(17) of \cite{psalt} or eq.(52) of \cite{psco}). A significant difference in  the two equations arises  in form of the contribution from  the repulsion part of drift term;  contrary to non-chiral case with repulsion arising from the terms of type $\frac{1}{e_n- e_m}$, the chiral case also contains additional terms of type $\frac{1}{e_n + e_m}$  and $\frac{(\nu+1/2)}{ e_n}$ (the terms arising due to existence of equal and opposite pairs of eigenvalues as well as zero eigenvalues and their repulsions). The additional terms are however relatively negligible for the spectral ranges $|e| \gg 0$ and eq.(\ref{ch19}) can be approximately reduce to the non-chiral case. This suggests analogous  statistical behavior, away from $|e|=0$, for the two cases which is also numerically confirmed by previous studies.

\subsection{Fluctuation measures}

The $Y$-based formulation  of the spectral JPDF, given by eq.(\ref{ch19}), indicates its applicability to a wide range of chiral ensembles. It also leads to the diffusion equation for $n^{th}$ order spectral correlation $R_n(e_1,...,e_n)$  i.e the probability density of $n$ positive eigenvalues irrespective of the location of  other ones, defined as $R_n(e_1,...,e_n) = {N! \over (N-n)!} \; \int \prod_{j=n+1}^N \; {\rm d}e_j \; P_e(e_1,...e_N)$. Using the spectral density formula $\rho(e)= \sum_n \delta(e-e_n)$, $R_n$ can also be expressed as the $n^{\rm th}$ order ensemble averaged level-density correlation: $R_n(e_1,...,e_n) = \langle \; \rho(e_1) \; \rho(e_2)...\; \rho(e_n) \; \rangle$ with $\langle . \rangle$ implying the ensemble average. Thus $R_n(e_1,...,e_n)$, for $n >1$,  describe the local fluctuations of the spectral density $\rho(e)$ and $R_1$ as its ensemble average. 

 The spectral density in general is system-dependent. 
For comparison of  the local fluctuations imposed on different spectral density backgrounds,  it is imperative to  rescale or  ''unfold'' the levels $e_j$ by the local mean level spacing $\Delta_{loc}(e)$ at spectral point of interest, say $e$. The rescaled correlations can be given as ${\mathcal R}_n(r_1,...,r_n; e) =  {\rm Lim} N \rightarrow \infty \; {R_n(e_1,...e_N) \over R_1(e_1)....R_1(e_n)} $ with $r_n={(e_n-e)\over \Delta_{loc}(e)}$ and $r=e \; \Delta_{loc}(e)$.  
The unfolding of eigenvalues  however also rescales the parameter $Y$; the rescaled complexity parameter can be given as \cite{chbe}
\begin{eqnarray}
\Lambda_e(Y,e)={ (Y-Y_0) \over (\Delta_{loc}(e))^2}.
\label{alm1}
\end{eqnarray}
As clear from the above, the rescaling  results in an energy dependence of the diffusion parameter.  

The diffusion equation for ${\mathcal R}_n$ for chiral Brownian ensemble is discussed in \cite{chbe} and is applicable for the present case too  (due to analogy of the diffusion equations for $P_N$ for the two cases) but with $Y$ now  given by eq.(\ref{y}). As discussed in \cite{chbe},  $\Lambda_e$ is obtained by neglecting the variations of  the average level density at $e$ \cite{chbe} and is therefore applicable for  the $n^{th}$ order local correlations within an energy range, say $e \pm n \Delta_{loc} e$,  with $\Delta_{loc}(e)$ dependent on the interaction among states.

\vspace{0.1in}

{\it Local mean level spacing:} A determination of $\Lambda_e$  from eq.(\ref{alm1})  requires a prior knowledge of   $Y-Y_0$ as well as $\Delta_{loc}$.  While $Y$ is explicitly given by eq.(\ref{y}),  $\Delta_{loc}$ depends on the underlying eigenfunction dynamics; for the eigenfunctions with non-ergodic dynamics, it can be significantly different from the mean level spacing $\Delta(e)$. This can be explained as follows: while $\Delta_{loc}(e)$  corresponds to  only those states at energy $e$ which are interacting, $\Delta(e)$ refers to all states at energy $e$ irrespective of their interaction. As the  eigenfunctions localized in different parts of the basis space do not interact, $\Delta_{loc}(e)$ is intuitively expected to be proportional to the average correlation/ localization volume at energy $e$. Based on the above reasoning, one possible definition can be given as follows: $\Delta_{loc}(e)={1\over \langle \rho_{loc} \rangle}$ where
\begin{eqnarray}
\rho_{loc}(e) = \sum_n \phi_n \; \delta(e-e_n)
\label{roc}
\end{eqnarray}
with $\phi_n$ as the probability of $n^{th}$ eigenfunction interacting with other eigenfunctions at energy $e$ (and therefore occupying the same region as other eigenfunctions with energies close to $e$).  
(Note $\rho_{loc}(e)$ here is not the standard local density of states which refers to the average spectral density at a specific basis location and energy; instead it refers to, in our case, number of interacting states at energy $e$ (and therefore, to the probability of states, with their energy $~e$, occupying the same region). For  example, for the case  in which a typical eigenfunction is delocalized in entire Hilbert space (e.g. a GOE), the above implies $\phi_n = 1$ which gives $\langle \rho_{loc}(e)\rangle=R_1(e)$ and $\Delta_{loc}(e)=\Delta(e)$. (Note, a GOE has a strong level repulsion and  a semicircle level density: $R_1(e) = \sqrt{2N -e^2}$. The latter is almost constant for a large neighborhood of $e \sim 0$ if $N$ is large; this  in turn implies $\langle \rho_{loc}(e)\rangle=R_1(e)$).

Similarly, in the case of localized dynamics e.g , although  two localized states do not typically overlap but can be localized in the same region with a small probability of $\xi^d/(2N)$ (with $\xi(e)$ as the average localization radius at energy $e$, $d$ as the system-dimension and $2N$ as the number of basis states) \cite{cue}. This implies $\phi_n \sim {\xi^d \over 2N}$ which gives $\rho_{loc}(e) = {\xi^d \over 2N} \; R_1(e)$. Using $R_1(e) = {1\over \Delta(e)}$, this leads to 
\begin{eqnarray}
\Delta_{loc}(e) = \Delta(e) \; {2N \over \xi^d}
\label{dele}
\end{eqnarray}
(with $\xi^d/2N$ as the probability of eigenfunctions localized in the same region of basis space). 
For cases where the eigenfunctions are exponentially localized e.g. in standard Anderson Hamiltonian (a single particle moving in a random potential), $\xi^d$ can be approximated by the average inverse participation ratio $\langle I_2 \rangle$ of the eigenfunctions with energies $\sim e$: $\xi^d \approx (\langle I_2 \rangle)^{-1}$. The latter relation however is not valid in general and one has to use the 
alternate routes to determine $\xi$. 

As discussed in \cite{chbe}, 
$\Lambda_e$ is the only parameter (besides energy range of interest) which appears in the differential equations determining the local spectral fluctuations. The latter are therefore expected to be analogous for two different ensembles if (i) both have same $\Lambda_e$ value, and, (ii) both evolve from an analogous initial condition (statistically). (Note, as mentioned below eq.(\ref{chi8}), the final end point of the diffusion is  a chiral GOE/GUE with the matrix element variance dependent on $\gamma$).  Equivalently, the ensembles with different  system conditions but subjected to same global constraints (e.g  Hermitian as well chiral nature of $H$-matrix in the present study) statistically corresponds to different crossover points on a specific curve (based on the global constraints) lying between the initial point $Y_0$ and the end point Ch-GOE/Ch-GUE.

\section{Numerical Verification of single parametric formulation }

To verify the above prediction, we numerically  compare the spectral fluctuations of four multi-parametric Gaussian ensembles of real-symmetric as well as complex Hermitian chiral matrices with different variance types. The ensemble-details  needed to determine  
$Y-Y_0$ as well as $\Delta_e$  and thereby $\Lambda_e$ are discussed below. The section also illustrates as to how different ensembles, if subjected to same matrix constraints, can be justified to evolve from same initial condition.

\subsection{Details of the Ensembles}

  The ensembles can briefly be described as follows.

\vspace{0.1in}

{\bf Chiral Anderson Ensemble (Ch-AE):} 
Within tight-binding approximation, the Hamiltonian $H$ of a $d$-dimensional bipartite lattice with ${\mathcal N}$ unit cells, each consisting of $2$ atoms,  with a single orbital contributing for each atom, can be given as 
 \begin{eqnarray}
H = \sum_{x,y}   V_{xy} \; c_x^{\dagger} . c_y 
\label{h1}
\end{eqnarray}
with   
$c^{\dagger}_x, c_x$ as the particle creation and annihilation operators on the site  $x$ with 
$V_{xx}$ as the on-site energy and $V_{xy}$ as the  hopping between sites $x,y$. Here $x=(m, \alpha)$ with $m$ as the label for the $d$-dimensional unit cell, $\alpha$ is the atomic label, $\alpha=a,b$. The motivation for choice of this system comes from the rich physical properties it has been shown to display by previous studies \cite{psf}.

To preserve chiral symmetry, here we consider the case with zero diagonal disorder \cite{ek},  a Gaussian hopping between atoms within a same unit cell and an isotropic Gaussian hopping between $z$ nearest neighbours sites on different unit cells; this implies (i) $ V_{xx}=0$, (ii) $V_{xy}\not=0$ and Gaussian distributed for $x=(m,a), y= (m, b)$, (iii) $ V_{xy} \not=0$ and Gaussian distributed  if $x=(m, \alpha)$ and $y= (m-1, \beta)$ or  $(m+1, \beta) $ with $\beta=a, b$ but $\beta\not=\alpha$, (iv) $V_{xy}=0$ for all other $x, y$ pairs.

In site basis,  the condition (i) results in a chiral structure  of matrix $H$ in eq.(\ref{ch1}) 
The conditions (ii) and (iii) lead to non-zero $C_{kk}$ and $C_{kl}$, respectively, for $k,l$ pairs corresponding to $z$ nearest neighbours sites on different unit cells. The ensemble density in this case is given by (\ref{rhoc}) with

\begin{eqnarray}
h_{kk} = \langle C_{kk}^2\rangle = {w^2/12}, \qquad 
h_{kl} =  \langle C_{kl}^2\rangle = f_1 \; {w_s^2/ 12}, \qquad b_{kl} =\langle C_{kl} \rangle = f_2 \; t
\label{vand}
\end{eqnarray}
where $f_1(kl)=1, f_2(k,l)=1$ for $\{k,l\}$ pairs representing hopping,   $f_1(k,l), f_2(k,l)\rightarrow 0$ for all $\{k,l\}$ values corresponding to disconnected sites. 
From eq.(\ref{y}), the ensemble complexity parameter in this case is \cite{psand}
\begin{eqnarray}
Y=-\frac{\beta N}{2M\gamma} {\rm ln} \left[|1-\gamma w^2/6| |1-\gamma w_s^2/6|^{z} \; |t+\delta_{t0}|^{2z} \right] + c_0
\label{y8}
\end{eqnarray}
where $M=\beta N (N+2 z) $ with  $2 \beta z N$ as the number of nearest neighbours which depends on the lattice conditions as well as the dimensionality $d$ of the system. Here $c_0$ is a constant of integration (determined by the initial condition on the ensemble).

To determine $Y_0$, the initial state is chosen as a clean bipartite lattice with sufficiently far off atoms resulting in zero hopping (i.e both $w=w_s=0$); consequently the initial ensemble corresponds to a localized eigenfunction dynamics with Poisson spectral statistics (with chiral constraint) and $Y_0=c_0$. Substitution of eq.(\ref{y8}) in eq.(\ref{alm1}) with $\Delta_{loc}(e) = {2N \langle I_2 \rangle \over R_1 }$ and $\overline{\langle I_2 \rangle}$ as the typical ensemble as well as spectral averaged inverse participation ratio (IPR) at $e$, leads to  the spectral complexity parameter 

\begin{eqnarray}
\Lambda_{e, A}(Y,N, e)   = {R_1^2 \over 8 \gamma  N^3 \; \overline{\langle I_2 \rangle}^2} \;  \ln \left[|1- \gamma \; w^2/6| \;  |1-\gamma \; w_s^2/6|^{z} \;|t+\delta_{t0}|^{2z}\right].
\label{alma}
\end{eqnarray}

\vspace{0.1in}

{\bf Chiral Rosenzweig-Porter Ensemble (Ch-RPE):}  This is a chiral variant of the standard Rosenzweig-Porter ensemble \cite{psrp, psbe}, with $\rho(H)$  given by eq.(\ref{rhoc2})  where
\begin{eqnarray}
  h_{kk;s}= \langle C_{kk;s}^2\rangle =1,  \quad  h_{kl;s} = \langle C_{kl;s}^2\rangle = \frac{ 1}{ (1+\mu)},\qquad k\neq l, \qquad b_{kl;s} = \langle C_{kl;s}\rangle = 0 \quad (\forall\; k,l) \nonumber \\
\label{ch23}
\end{eqnarray}
with $0<\mu <\infty$; the limits $\mu= 0$ and $\infty$ correspond to a chiral Gaussian orthogonal ensemble (Ch-GOE) and a matrix ensemble with diagonal chiral blocks, respectively.
Substitution of the above values in eq.(\ref{y}) gives $Y$ for this case
 
\begin{eqnarray}
Y = -\frac{\beta N(N-1)}{2M\gamma}{\rm ln} \left[1-\frac{2\gamma}{(1+\mu)}\right] + c_0
\label{ybe}
\end{eqnarray}
with $M=\beta N (N+\nu) = \beta N^2$ (with $\nu=0$ in our numerics) and $c_0$ as a constant of integration (determined by the initial condition on the ensemble). 

Choosing initial condition with $\mu \rightarrow \infty$ corresponds to  an ensemble of $H$ matrices,  with its chiral blocks as diagonal $C$-matrices and the spectral statistics as  Poisson statistics (with chiral constraint).  From eq.(\ref{ybe}, this implies $Y_0=c_0$ and $Y-Y_0 \approx {1\over  \mu}$ (for $\mu \gg 1$). This on substitution in eq.(\ref{alm1}) leads to the spectral complexity parameter for Ch-RPE case

\begin{eqnarray}
\Lambda_{e,B}(e)=\frac{Y-Y_0}{\Delta_{loc}(e)^2} \approx  \frac{R_1^2}{  \mu }.
\label{almb}
\end{eqnarray}
Here the 2nd equality  is obtained by using $\Delta_{loc}(e)=\Delta(e)= {1\over R_1(e)}$. As clear from the above,  $\Lambda_{e, B}$ depends on three   parameters, namely, $\mu$, matrix size $N$ as well as the spectral location $e$ chosen for the analysis of the local fluctuations.

The ensemble density with distribution parameters  given by eq.(\ref{ch23}) is analogous to a specific class of chiral Brownian ensemble (Ch-BE) \cite{chbe}, namely, that   arising due to a single parametric perturbation of a chiral ensemble, with Poisson statistics for non-zero eigenvalues,  by a Ch-GOE ensemble; we henceforth refer the Ch-RPE case as Ch-BE case.

\vspace{0.1in}

{\bf Chiral Gaussian ensemble with Power law decay (Ch-PE):}
The  $C$-matrix ensemble in this case consists of independently distributed Gaussian entries with  zero mean  and a power law decay of variances away from the  diagonal. The ensemble density  $\rho(H)$ can again be described by eq.(\ref{rhoc2}) with 
\begin{eqnarray}
h_{kl} =\langle C_{k,l}^2\rangle = \frac{1}{1+\frac{|k-l|^2}{b^2}},\qquad  b_{kl} =\langle C_{kl}\rangle =  0 \quad \forall \;\; k, l 
\label{ch25}
\end{eqnarray}
where $b$ is arbitrary parameter. Eq.(\ref{y}) then gives 
\begin{eqnarray}
Y=-\frac{\beta}{2M\gamma}\left[ \sum_{r=0}^{N} g_r \; (N-r) \; {\rm ln} \left(1-\frac{2\gamma}{1+(r/b)^2}\right)\right] + c_0
\label{ch26}
\end{eqnarray}
with number of independent elements $M=\beta N^2$, $r\equiv |k-l|$ and $g_r=(2-\delta_{r0})$. 
Here the case $b \ll 1$ corresponds to $H$-ensemble with diagonal $C$-matrices with Poisson spectral statistics and can therefore be chosen as the initial ensemble. The choice  leads to $Y_0=-\frac{N \beta}{2M\gamma} \; {\rm ln} \left(1-{2\gamma} \right)+c_0 \approx c_0$ (for large $N$).

The above along with eq.(\ref{alm1}), with $\Delta_{loc}(e) = {2N \langle I_2 \rangle \over R_1 }$ then leads to
\begin{eqnarray}
\Lambda_{e,P}(b,e) =  {R_1^2 \over 8 \gamma  N^4 \; \overline{\langle I_2 \rangle}^2} \; \left[ \sum_{r=1}^{N}  \; (N-r)\;{\rm ln} \left(1-\frac{2\gamma}{1+(r/b)^2}\right)^2\right] 
\label{almp}
\end{eqnarray}
 The spectral statistics of Ch-PE therefore shows a crossover from 
from Poisson  (for $\Lambda_{e,P}\rightarrow 0$ 
as $b \rightarrow 0$) to Chiral GOE behavior (for 
$\Lambda_{e,P}\rightarrow \infty$ as $b \rightarrow \infty$).

\vspace{0.1in}

{\bf Chiral Gaussian ensemble with exponential decay (Ch-EE): }
here the ensemble of $C$-matrices  corresponds to  an exponential decay of the variances 
away from the diagonals $C_{kk}$ but with mean $\langle C_{kl}\rangle = 0$ for all $k,l$. The $\rho(H)$ is again given by eq.(\ref{rhoc2}) with 
\begin{eqnarray}
h_{kl}=\langle C_{kl}^2\rangle = {\rm exp}\left(-\frac{|k-l|}{b}\right)^2,\qquad  b_{kl}=\langle C_{kl}\rangle = 0 \qquad  \forall \;\; k, l 
\label{ch27}
\end{eqnarray}
with $b$ as an arbitrary parameter. Eq.(\ref{y}) now gives  
\begin{eqnarray}
Y=-\frac{\beta }{2M\gamma}\left[ \sum_{r=0}^{N} g_r \; (N-r)\;{\rm ln} \left(1-\frac{2\gamma}{{\rm exp}(\frac{r}{b})^2}\right)\right]  + c_0
\label{ch28}
\end{eqnarray}
with $M=\beta N^2$, $r\equiv |k-l|$ and  $g_r=2-\delta_{r0}$.

To keep analogy with the other ensembles described above, here again  the initial ensemble for $C$ is chosen  that of the diagonal matrices with  a Poisson spectral statistics which  corresponds to  $Y_0  = -\frac{N \beta}{2M\gamma} \; {\rm ln} \left(1-{2\gamma}\right) +c_0 \approx c_0$ (for large $N$).  
Referring the localization length as $\xi$ and using $\Delta_{loc}(e) = {2 N \over \xi} \; \Delta(e)$, the spectral complexity parameter now becomes 
\begin{eqnarray}
\Lambda_{e,E}(b,e) =   \; \frac{\xi^2 R_1^2}{8 \gamma N^4}\left[\sum_{r=1}^{N}  \; (N-r)\;{\rm ln} \left(1-\frac{2\gamma}{{\rm exp}(\frac{r}{b})^2}\right)\right]
\label{alme}
\end{eqnarray}

 \vspace{0.1in}
 
\subsection{Numerical Analysis}

For numerical analysis of local spectral fluctuations for each case mentioned in  section IV.A, we exactly diagonalise  (using LAPACK, a  standard software library for numerical linear algebra subroutine for complex matrices   \cite{nlb}),  the ensembles for many system parameters   but with a fixed $\gamma=1/4$.  The $C$ matrix chosen for all cases considered here is a $N \times N$ square matrix which corresponds to $\nu=0$.  The other system-related details used in our numerics for each case  are as follows (also given in tables I-V):

(i) {\it Ch-AEs}: we consider the  ensembles of  $2N \times 2N$  matrix  $H$ (eq.(\ref{h1}) for two dimensional ($d=2$) bipartite square lattice  (of linear size $L$ with $L^2=2N$) subjected to periodic boundary condition; the ensembles parameters are given by eq.(\ref{vand}) with $z=4$. Variation of matrix size and disorder strength leads to four different ensembles: one consisting of $5000$ matrices of size $2N=1024$ and, another of $2500$ matrices of size $2N=2116$, each analyzed for two disorder strengths $w^2=12$ and $36$ (keeping $w_s^2=12$  and $t=0$ for both cases).

(ii) {\it Ch-RPE}: here we choose the ensemble described by eq.(\ref{ch23}) with $\mu=c \; N^2$; (the choice is motivated by previous studies of non-chiral Rosenzweig-Porter ensembles (or Brownian ensembles) which confirm this $\mu$-value as a critical point of the statistics \cite{psbe, psand}). The ensemble is exactly diagonalised for  two different $c$ values, i.e. $c=1$ and $c=0.4$, each case considered for two different ensembles: one consisting of 5000 matrices of  sizes $2N=1000$ and another of 2890 matrices of size $2N=1728$.

(iii) {\it Ch-PE}: The numerics in this case is considered for the ensemble (\ref{ch25}),  of $5000$ matrices of size $2N=1000$ and another of $2500$ matrices of size $2N=2000$; each ensemble is analyzed for  two $b$-values i.e $b=0.5$ and $0.75$.

(iv) {\it Ch-EE}: here again we consider an ensemble (\ref{ch27}) of $5000$ matrices of size $2N=1000$ and another of $2500$ of size $2N=2000$; both ensembles are  analyzed for  two $b$-values i.e $b^2=100$ and $144$.

 The local fluctuations of the spectral density of a  complex system are often imposed on a smooth background,   (i.e average spectral density), varying from one system to another.  It is necessary, for a meaningful comparison of the statistics, to  rescale the spectrum by the ensemble averaged level density $R_1(e)$ (referred as unfolding)  \cite{fh}.   Due to often unavailability of the analytical form of $R_1(e)$  for complex systems, the standard route is to determine it through  numerical calculation. 
 But for systems, whose  $R_1(e)$ is not a smooth function of energy, the unfolding procedure becomes nontrivial even if $R_1(e)$ is analytically known and the spectrum is stationary. Further, in case of non-stationary spectrums,  there are additional complications; this is because the fluctuations remain energy-dependent even after unfolding (as indicated by the energy-dependence of $\Lambda_e(e)$) \cite{chbe, psco, psand}.  For comparison of local statistics, therefore, ideally one should consider an ensemble averaged fluctuation measure at a specific energy-point, say $e$, without any spectral averaging. This in general requires consideration of huge ensembles and runs into practical technical issues. Fortunately, in the spectral regions  where $\Delta_{loc}$ varies very slowly, it is  possible to choose an optimized range  in the neighborhood of $e$, sufficiently large for good statistics but keeping  a mixing of different statistics at minimum. This is however not the case for the regions with sharp change of $\Delta_{loc}$; the latter leads to a rapidly changing  $\Lambda_e$  and it is numerically difficult to consider a spectral range with an appropriate number of levels  without mixing of different statistics. This compels us to consider, for numerical analysis,   only ${1 \% }$ of the total eigenvalues taken from a range $\Delta_{loc}(e)$ around $e$ if $e$ is in the bulk.  A rapid variation of the spectral density (e.g near $e=0$ or spectral-edge) however permits one to consider a very small spectral ranges ($~0.5 -1 \%$); this in turn  gives rise to errors in $\Lambda_e$ calculations (as evident from figures 7 and 8).

 Almost all standard spectral fluctuation measures e.g  nearest neighbour spacing distribution and number variance are sensitive to unfolding issues which can not be ignored especially in case of a non-stationary spectrum. This motivated the study \cite{huse} to  introduce a new measure, namely, the nearest neighbour spacing ratio distribution $P(r)=\sum_{i=1}^{N-1}\langle \delta(r-r_i)\rangle$ with  $r$ defined as the ratio of consecutive spacings between nearest neighbor levels: $r_i=s_{i+1}/s_i$ where $s_i=e_{i+1}-e_i$ is the distance between two nearest neighbour eigenvalues \cite{huse, abgr}. As the ratio $r$ does not depend on the local density of states,  an  unfolding of the spectrum for $P(r)$ is not required \cite{huse}. Further $P(r)$ being a short range fluctuation measure,  it reduces the chances of mixing spectral statistics. For the spectral statistics in Poisson and Wigner-Dyson limit, $P(r)$ can be given as \cite{abgr}
 \begin{eqnarray}
 P(r) &=& \frac{c_{\beta} \;  (r+r^2)^{\beta} }{(1+r+r^2)^{1+(3/2)\beta}} \qquad {\rm Wigner-Dyson}  \label{wd} \\
 &=& \frac{1}{(1+r)^2}  \hspace{1.3in} {\rm Poisson}
 \label{po}
 \end{eqnarray}
 with $c_1=\frac{27}{8}$ and $c_2=\frac{81 \sqrt{3}}{4\pi}$. 
 
 In the regime intermediate to Poisson and GOE/GUE, our theory suggests  $P(r)$ to be governed only by $\Lambda_e$. A recent study \cite{relan} has indeed postulated  a one-parameter distribution for $P(r)$ in the intermediate regime 
 \begin{eqnarray}
P(r;\beta_t,\gamma(\beta_t))=C_{\beta_t} \;\frac{(r+r^2)^{\beta_t}}{\left[(1+r)^2-\alpha(\beta_t)\;r\right]^{1+1.5 {\beta_t}}}
\label{pr}
\end{eqnarray}
with $C_{\beta_t}$ as a normalization constant defined by the condition $\int_0^{\infty} {\rm d}r \; P(r)=1$. Here $\alpha(\beta_t)$ is  defined by the ideas based on information entropy \cite{relan}:
with  $\alpha(\beta_t)=0.92-1.42\;(2-\beta_t) +0.01\;(2-\beta_t)^7$ (for $\beta=2$) and 
$\alpha(\beta_t)=0.80-1.69\;(1-\beta_t) +0.89\;(1-\beta_t)^5$ (for $\beta=1$) with $\beta_t$ as the fitting parameter: $0 \le \beta_t \le \beta$. The desire to understand the connection between our $\Lambda_e$ and $\alpha$ in eq.(\ref{pr}) led us to fit our numerical results for $P(r)$ with eq.(\ref{pr}); our analysis suggests a linear relation between them (see tables I-VI).

From  eq.(\ref{alm1}), $\Lambda_e$ for an ensemble can be determined if  $R_1(e)$ as well as ensemble averaged  localization length $\xi$ is known. The latter can often be estimated (e.g. for Ch-AEs and Ch-PEs) from  the average inverse participation ratio  $\langle I_2 \rangle$. The theoretical formulations  for $R_1(e)$ and $\langle I_2 \rangle$  for the cases used in our numerics  are however not  known. (It is worth emphasizing here that such information is in general not available for most of the multi-parametric ensembles especially those with sparse matrix structures). Although, for non-chiral BE case, with $\mu=c N^2$, $R_1(e)$ is theoretically known to be a Gaussian \cite{shep} but its validity for Ch-BE is not a priori obvious. Further,  while the average level density of chiral Anderson Hamiltonian is discussed in previous studies (see for example \cite{ek} and references therein), it is not exact and also  based on the numerical analysis for specific system parameters.
In absence of a theoretical formulation, the option left to us is to determine $R_1(e)$ and $\xi$  by a numerical analysis. For Ch-EE-case, however we find that the approximation $\xi \sim \langle I_2 \rangle^{-1}$ does not seem to be valid; instead using $\xi \approx \sqrt{2 N}$ (following insight based on numerics) gives  results consistent with our theoretical claim  about $\Lambda_e$.

Figure 1 illustrates the energy as well as system dependence of the scaled level-density $F(e)=R_1(e)/2N$  for the ensembles (i)-(iv). As clear from the figures 1.(a), 1.(c) and 1.(d), Ch-AE, Ch-PE and Ch-EE show a strong dependence of $F(e)$ on the variances $h_{kl}$ of the matrix elements $C_{kl}$  but insensitivity to the  matrix size $N$.  On the contrary,  part (b) for Ch-BE  indicates the independence of $F(e)$ from both $h_{kl}$ as well as $N$.  An important point worth noting here is a weakly singular level density near $e=0$ in figure 1.(a) (although $\nu=0$ for our case); the behavior arises due to choice of zero mean off-diagonal randomness  in $C$-matrix (for its non-zero elements) and is consistent with previous studies \cite{ek}. A weaker singularity  displayed in figures 1.(c,d) is a result of weaker relative sparsity of the $C$ matrix elements in case of Ch-PE and and Ch-EE. The absence of singularity in figure 1.(b) results from the lack of sparsity in $C$-matrix of Ch-BE; note the Gaussian form in figure 1.(b) is consistent with theoretical prediction of \cite{shep}; (although the latter study is on non-chiral BEs).

Figure  2 illustrates  the energy as well as system dependence $\langle I_2 \rangle$ for the four ensembles. As the parts (a) and (c) indicate, $\langle I_2\rangle$ for Ch-AE and Ch-PE is sensitive to the variance of the matrix elements but not to the matrix sizes $N$.
Although $\langle I_2 \rangle$   does not appear in our $\Lambda_e$ formulation for Ch-BE and Ch-EE, its behavior for these cases is  still displayed in parts (b) and (d) of the figure 2 for comparison with other cases.

Using $F(e)$ and $\langle I_{2}\rangle$ (latter used only for Ch-AE and Ch-PE) at a given $e$ from figures 1 and 2,  we calculate $\Lambda_e(e)$ for the four ensembles. 
As eqs.(\ref{almb}, \ref{alma}, \ref{almp}, \ref{alme}) indicate,  $\Lambda_e$ for each ensemble not only depends on the energy-range $e$  of interest but also on at least two other system-parameters. The  analogy of the local spectral statistics among the ensembles can then manifest in many ways. More clearly, if indeed governed only by $\Lambda_e$ as predicted by our theory, an analogy for the local correlations at different energies can occur within the same ensemble but by varying other distribution parameters. For example if $\Lambda_{e,x}(e_1, s_1)=\Lambda_{e,x}(e_2, s_2)$ with  $s_1, s_2$ referring to two different sets of system parameters for a specific ensemble $''x''$ (e.g $x=A, B, P, E$), the local correlations of the latter at $e_1, s_1$ are expected to be analogous to those at $e_2,s_2$; this is later on referred as the ''intra-system'' analogy.  Similarly if $\Lambda_e$ for different ensembles are equal (for same or different $e$-values), their local correlations  should be analogous too. For example if $\Lambda_{e,x}(e_1)=\Lambda_{e,y}(e_2)$ with $x, y=A, B, P, E$, the local correlations for the ensemble $x$ at $e_1$ are then predicted to be analogous to that of $y$ at $e_2$; (later referred as the ''inter-system'' analogy).

An important point to note here is the following: for $\Lambda \to 0, \infty$ (stationary limits), $P(r)$ is expected to approach Poisson and GOE/ GUE limits, respectively, for almost all $e$ ranges and becomes $e$-independent. A variation of the statistics from Poisson to GOE/GUE limits, as $e$ varies, occurs only for finite, non-zero $\Lambda_e(e)$; (the latter corresponds to a critical regime of the statistics for finite $N$ and a critical point in large $N$-limits). The condition $\Lambda_{e,x}(e_1, s_1)=\Lambda_{e,x}(e_2, s_2)$ can therefore be satisfied only if $\Lambda_e$ remains finite, non-zero  as well as $e$-dependent for both cases.

To confirm our theoretical prediction, here we numerically verify both these analogies by comparing $P(r)$  for a number of combinations. The details are as follows. 

\vspace{0.1in}

{\bf Intra-system analogy:} This concerns with the local spectral statistics of the ensembles consisting of a same Hamiltonain matrix representing a system, say ''x''. The ensemble parameters  for the analogs  can then be obtained by  invoking following condition 

\begin{eqnarray}
\Lambda_{e,x}(e_1,s_1)=\Lambda_{e,x}(e_2, s_2)= \Lambda_{e,x}(e_3, s_3)=\Lambda_{e,x}(e_4, s_4).
\label{almu}
\end{eqnarray}
with $\Lambda_{e, x}$ is given by eq.(\ref{alma}), eq.(\ref{almb}), eq.(\ref{almp}) and eq.(\ref{alme}) for $x=A,B, P,E$ respectively. As mentioned above, the ensembles chosen should be non-stationary The above analogy being sensitive to error in $\Lambda$-calculation,  we avoid mixing of the statistics by taking only $1\%$ of the eigenvalues from the chosen energy range if $|e| >0$; the percentage of levels considered for $e \sim 0$ however is less (between $0.5-1 \%$).  

For  comparisons of the ensembles in different energy regimes, it is preferable to choose the same number of levels for each case.  For this purpose, the number of matrices $M$ in the ensemble for each matrix size $N$ is so chosen so as  to give approximately $2.5 \times 10^4$ eigenvalues  for the analysis.  Further as eqs.(\ref{y8}, \ref{ybe}, \ref{ch23}, \ref{ch26}) indicate, $Y$ is $\beta$-independent. Thus $\Lambda_e$  depends on $\beta$ only through $\Delta_{loc}$ (more specifically through average localization length $\xi$). The effect however is quantitative only and does  not lead to any qualitatively new insights in case of intra-system analogy. To avoid repetition, here we  consider $\beta=1$ case only.

Figure  3 displays a comparison of $P(r)$ behavior for  Ch-AE obtained from four different combinations of disorder $w$ and system size $N$ at  a specific energy $e$. Here the parametric combinations are chosen such that  eq.(\ref{almu}) is satisfied (with $\gamma=1/4$). (The numerical procedure used for the purpose is as follows: we arbitrarily choose a set of system parameters, say $s_1$, numerically obtain corresponding Ch-AE spectrum, say "AE1",  by exact diagonalization and  its $P(r)$ at an arbitrary spectral value, say $e=e_1$,  find $R_1(e)$ and $\langle I_2 \rangle$ at $e_1$ from figures 1 and 2 and substitute them in eq.(\ref{alma}) to obtain  $\Lambda_e(e_1, s_1)$.   The same procedure is then repeated to generate Ch-AE spectrum, say "AE2" for the system parameters set $s_2$ again chosen arbitrarily. But the $P(r)$ for "AE2" is numerically considered  at an $e=e_2$ value (taking levels within $\le 1\%$ of $e_2$) such that the equality $\Lambda_e(e_1,s_1)=\Lambda_e(e_2,s_2)$ is ensured.  Note the latter condition limits  the choice of $s_2$; it is arbitrary only to the extent that the relation $\Lambda_e(e_1,s_1)=\Lambda_e(e_2,s_2)$ can be satisfied within available spectral range for the system. Same procedure is then repeated for the Ch-AEs with parameters $s_3$ and $s_4$. 

As mentioned above, a variation of bulk-statistics between two end-points can only be seen if the ensemble is  non-stationary. In infinite $N$-limit, the non-stationarity in AEs occurs only at some critical system parameter (e.g. critical disorder) or spectral point (e.g.mobility edge) \cite{garcia, psand}, implying very few non-zero, finite $\Lambda_e$ values which leaves very few options for satisfying eq.(\ref{almu}). 
For finite system sizes however, the AEs are known to have a critical regime of statistics different from Poisson and GOE/GUE and one can analyze the validity of eq.(\ref{almu}) for many $\Lambda_e$ values; here we consider the comparison  for four  different $\Lambda_e$ values.   To indicate that the statistics is indeed changing with $\Lambda_e$, the two stationary limits, namely, $P(r)$ for Poisson and GOE cases (eqs.(\ref{wd}, \ref{po})) are also displayed in each part of the figure.

  The system-parameters for four Ch-AEs as well as the spectral parameters (i.e $e, R_1(e), \langle I_2 \rangle$),  used in figure 3  satisfying eq.(\ref{almu}),   are given in Table 1. The $8^{th}$ column of the latter gives the $\Lambda_e$ values calculated from eq.(\ref{alma}); here a small deviation can be seen due to unavailability of exact analytical form of $R_1(e)$ and $\langle I_2(e)\rangle$, needed to invert $\Lambda_e(e_1,s_1)$ to find exact system parameters for other AEs. (Note the column 1 of table I mentions only the approximate $\Lambda_e$ used as a label in the figure 3).

In contrast to Ch-AEs, with many system parameters, Ch-BEs, CH-PEs and Ch-EEs depend only on one parameter besides matrix size $N$ (see section IV.A). 
Figures 4-6 show $P(r)$ for four different Ch-BEs (obtained by changing $\mu,N,e$), Ch-PEs (different combinations of $b$ and size $N$ and $e$) and Ch-EEs (different combinations of $b^2, N, e$) respectively. 
Here again the $P(r)$-analogies are obtained by repeating the same procedure as for Ch-AEs mentioned above (now $\Lambda_e$ in eq.(\ref{almu}) given by  eqs.(\ref{almb}, \ref{almp}, \ref{alme}) for Ch-BE, Ch-PE and Ch-EE respectively). The corresponding parametric values and spectral ranges leading to almost same $\Lambda_e$ values   are given in Tables II, III and IV, (for figures 4, 5, 6, respectively).

Note, for smaller $\Lambda$-values, the figures 3-6 may seem to indicate a different rate of crossover from Poisson to GOE. This may seem to suggest a violation of our theory which predicts  statistical analogy if $\Lambda_e$ is same. We believe however this is due to numerical errors originating from (i)  spectral averaging used in the numerics, (ii) lack of exact information about $R_1(e)$ and $\Delta_{loc}(e)$. Also note $\Lambda_e(e)$ calculation is quite prone to errors for regions where $R_(e)$ changes rapidly e.g $e \sim 0$ as well as in the  spectral-edge region and near inflection points for Ch-AE and Ch-PE (see figure 1). 

Although not displayed in the figures 3-6, we also fit $P(r)$ for each $\Lambda$ with eq.(\ref{pr}) with $\beta_t$ and $C_t$ as the fitting parameters and find  $\alpha(\beta_t)=\Lambda-1.69\;(1-\beta_t) +0.89\;(1-\beta_t)^5$. The fitting parameters $\beta_t$ and $C_t$ for each case are given in tables I-IV.

\vspace{0.1in}

{\bf Inter-system analogy:}  In contrast to Ch-AEs, with many system parameters, Ch-BEs, CH-PEs and Ch-EEs depend only on one parameter (besides matrix size $N$). It is therefore natural to query whether eq.(\ref{almu1}) can successfully be used to map their statistics onto each other.  
The condition for the ensemble and spectral parameters  leading to the analogs  now becomes 

\begin{eqnarray}
\Lambda_{e, A}(s_1,e_1)=\Lambda_{e, B}(s_2,e_2) = \Lambda_{e, P}(s_3,e_3)= \Lambda_{e, E}(s_4,e_4)
\label{almu1}
\end{eqnarray}
with $\Lambda_{e, B}$, $\Lambda_{e, A}, \Lambda_{e, P},  \Lambda_{e, E}$ given by eq.(\ref{almb}), eq.(\ref{alma}), eq.(\ref{almp}) and eq.(\ref{alme})  respectively (with $\gamma=1/4$). Here again the analogy is obtained by varying  $e$-values; (note the analogy can also be studied for same $e$ values by a careful choice of ensemble parameters). As illustrated in Fig. 7, $P(r)$ for all the four cases with $\beta=1$ overlap with each other if their $\Lambda_e$ are equal. The details of  each  ensemble used in numerics for figure 7 are given in Table V.  Although the overlap seems to be poor for small $\Lambda$ values, this is again is due to numerical errors 
associated with $\Delta_{loc}(e)$ estimation; (in contrast to intra-system analogy, the error in  the latter becomes crucial for comparisons of different systems). These errors can be reduced if an exact  theoretical formulation of $\xi$ is available for the system of interest; (note although $\xi \sim \langle I_2 \rangle^{-1}$ seems to work for Ch-AE and Ch-PE but it is not an exact relation. Further, although  intuitive reasoning in section III.B gives $\Delta_{loc}(e)=\Delta(e)$ for RPE and  $\Delta_{loc}(e)=\sqrt{2N}\;\Delta(e)$ for Ch-EE is supported numerically,  an exact formulation of  $\Delta_{loc}(e)$ for RPE and  Ch-EE is still missing).

We consider the inter-system analogy for $\beta=2$ case too. The results are displayed in figure 8 along with Poisson and GUE limits; the  parametric details for this figure are given in table VI. (Although the figures for $R_1(e)$ and $\langle I_2(e) \rangle$ for $\beta=2$ case are not included in this work, but their  values used for $\Lambda_e$-calculation are given in table VI).  The figure reconfirms our theoretical prediction  regarding the insensitivity of local spectral statistics to specific system details and the role played by $\Lambda_e$.

We fit $P(r)$ with eq.(\ref{pr})  for all cases displayed in figures 7 and 8 and again find  $\alpha(\beta_t)=\Lambda-1.69\;(1-\beta_t) +0.89\;(1-\beta_t)^5$ for $\beta=1$ and $\alpha(\beta_t)=\Lambda -1.42\;(2-\beta_t) +0.01\;(2-\beta_t)^7 +0.01\;(2-\beta_t)^6$ for $\beta=2$. The  fitting parameters $\beta_t$ and $C_t$  for each case are given in tables V and VI.

\section{Conclusion}

In the end, we summarize with a brief discussion of our main results and open questions. Extending  the  complexity parameter formulation for Hermitian ensembles without chirality to those with chirality, we analyze, for the latter,   the statistical response  of the eigenvalues  to  multi-parametric variations and its reduction to the complexity parameter formulation. As the chiral cases include non-Hermitian matrix blocks, this not only renders  the technical analysis more complicated but also leads to the diffusion equations for the spectral JPDF different from the case without chirality.  But, as in the non-chiral case, the spectral complexity parameter in  the chiral case is again a function of  energy as well as ensemble parameters. This predicts an important connection hidden underneath the local spectral fluctuations of a complex system: its  statistics at an energy for a given set of system condition can be analogous to that at another energy but with different  system conditions.  For two different complex systems,  however, the analogy can occur even at same energy if their complexity parameters are equal and both belong to same global constraint class (i.e same symmetry class and conservation laws). Our  theoretical predictions are confirmed  by a numerical comparison of the spectral statistics  of four multi-parametric Gaussian ensembles with different sets of ensemble parameters and  at different energies. 

Although, in the present work, we have confined our analysis to spacing ratio distributions only (so as to minimise any error dus to unfolding issues of the spectrum), previous studies on non-chiral ensembles have analyzed other measures too e.g. nearest neighbor spacing distribution, number variance etc \cite{psand, pssymp, pswf, pskr, psf}, wave-function statistics \cite{pswf}, conductance distribution \cite{pscond} etc; these studies also support $\Lambda_e$ based universality of the local fluctuations.  As discussed in \cite{chbe}, the multi-parametric chiral ensembles are  connected to some other complex systems too e.g. the systems represented by multi-parametric Wishart ensembles and Calogero-Sutherland Hamiltonian (CSH). The results and insights in the statistics of any one of them have therefore important implications for the others. Further the appearance of Wishart matrices in wide ranging areas makes the results derived for Chiral ensembles useful for these areas too.  Considering not many theoretical results so far are available for system-dependent random matrix ensembles (e.g. sparse ensembles of disordered Hamiltonians), the theoretical predictions based on complexity parametric formulation  can be very useful and therefore need detailed investigations as well as experimental verifications if feasible.

Our study still leaves many questions unanswered. The first and foremost among them is a
theoretical formulation of $\Delta_{local}(e)$ used in eq.(\ref{alm1}). Although we have given an intuitive reasoning in section III.B, its exact formulation is still missing. Further the present work is confined only to spectral statistics, 
a similar comparison for the eigenfunction fluctuations  especially near zero energy is also very desirable (see \cite{chbe} for more details). Another important question is about the  transition from multi-parametric chiral ensembles to non-chiral ensembles as chiral symmetry is partially broken. We intend to pursue some of these studies elsewhere.

\vfill\eject

\vfill\eject

\begin{figure}[h!]
	\centering
	\includegraphics[width=1.2\textwidth,height=1.0\textwidth]{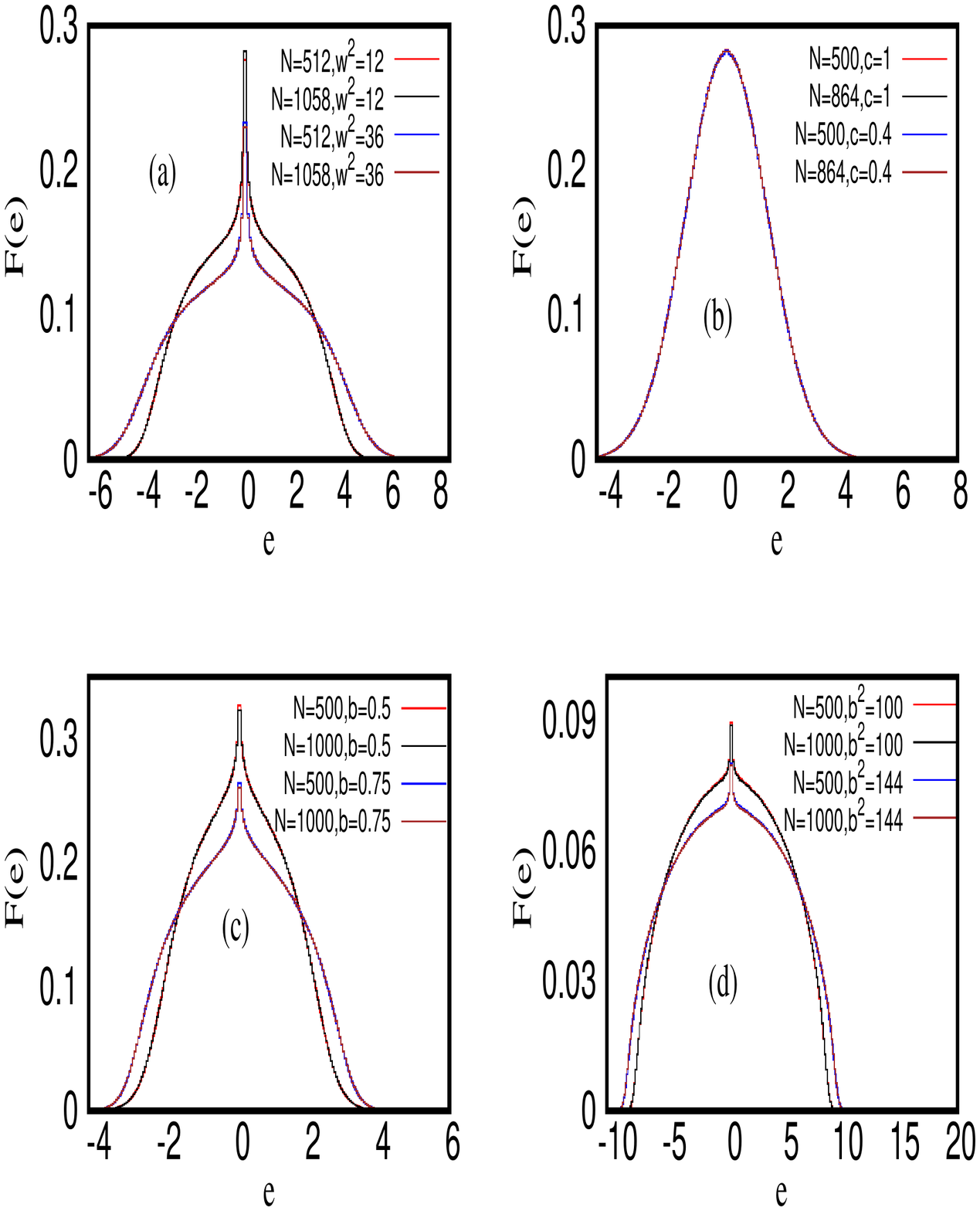}
	\caption{{\bf Density of states}: The determination of spectral complexity parameter $\Lambda_e$ requires a prior knowledge of the ensemble averaged level density. The parts (a),(b),(c), (d) display the scaled level-density $F(e) =R_1(e)/2N$  for the cases Ch-AE, Ch-BE, Ch-PE and Ch-EE respectively. As mentioned in the main text, $C$ chosen for our numerics is $N \times N$ square matrix thus implying $\nu=0$; note that however does not imply that the level density will  dip near $e=0$. As clear from the parts (a), (c) and (d), Ch-AE, Ch-PE and Ch-EE show a strong dependence of $F(e)$ on the variances $h_{kl}$ of the matrix elements but insensitivity to the matrix size $N$.  On the contrary,  part (b) for Ch-BE  indicates an independence of $F(e)$ from $h_{kl}$ as well as $N$.
}
\label{fig1}
\end{figure}

\begin{figure}[h!]
\centering
\includegraphics[width=1.2\textwidth, height=1.0\textwidth]{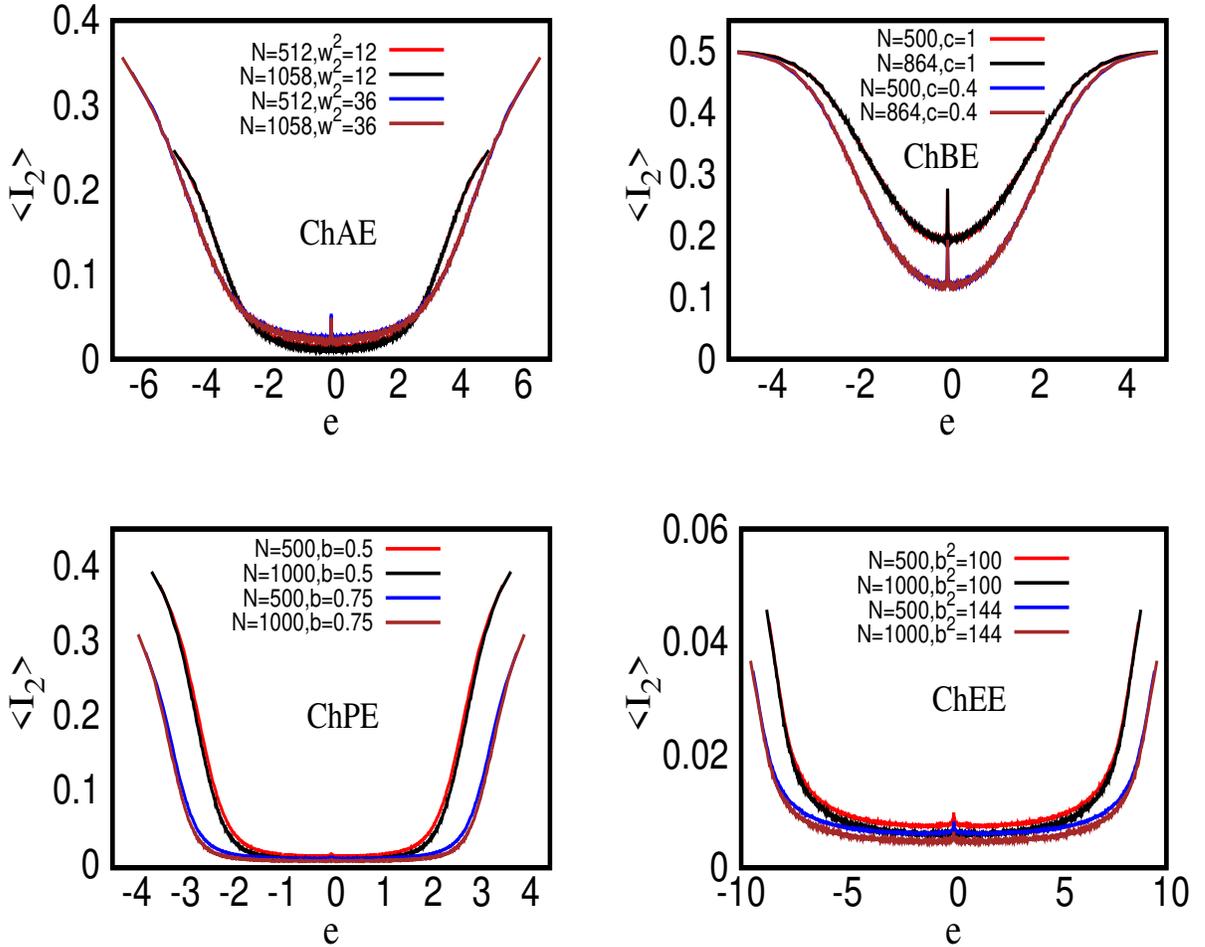}
\caption{{\bf Ensemble averaged inverse participation ratio (IPR)}:
As mentioned in the text, a prior knowledge of  $\langle I_2\rangle$ is needed to determine  $\Lambda_e$ for some cases ; the behavior for Ch-AE, Ch-BE, Ch-PE and Ch-EE is illustrated in parts(a),(b),(c),(d) respectively.
As clare from each part, $\langle I_2\rangle(e)$-behavior  is sensitive to the variance of the matrix elements.}
\label{fig2}
\end{figure}

\begin{figure}[h!]
	\centering
	\includegraphics[width=1.2\textwidth, height=1.0\textwidth]{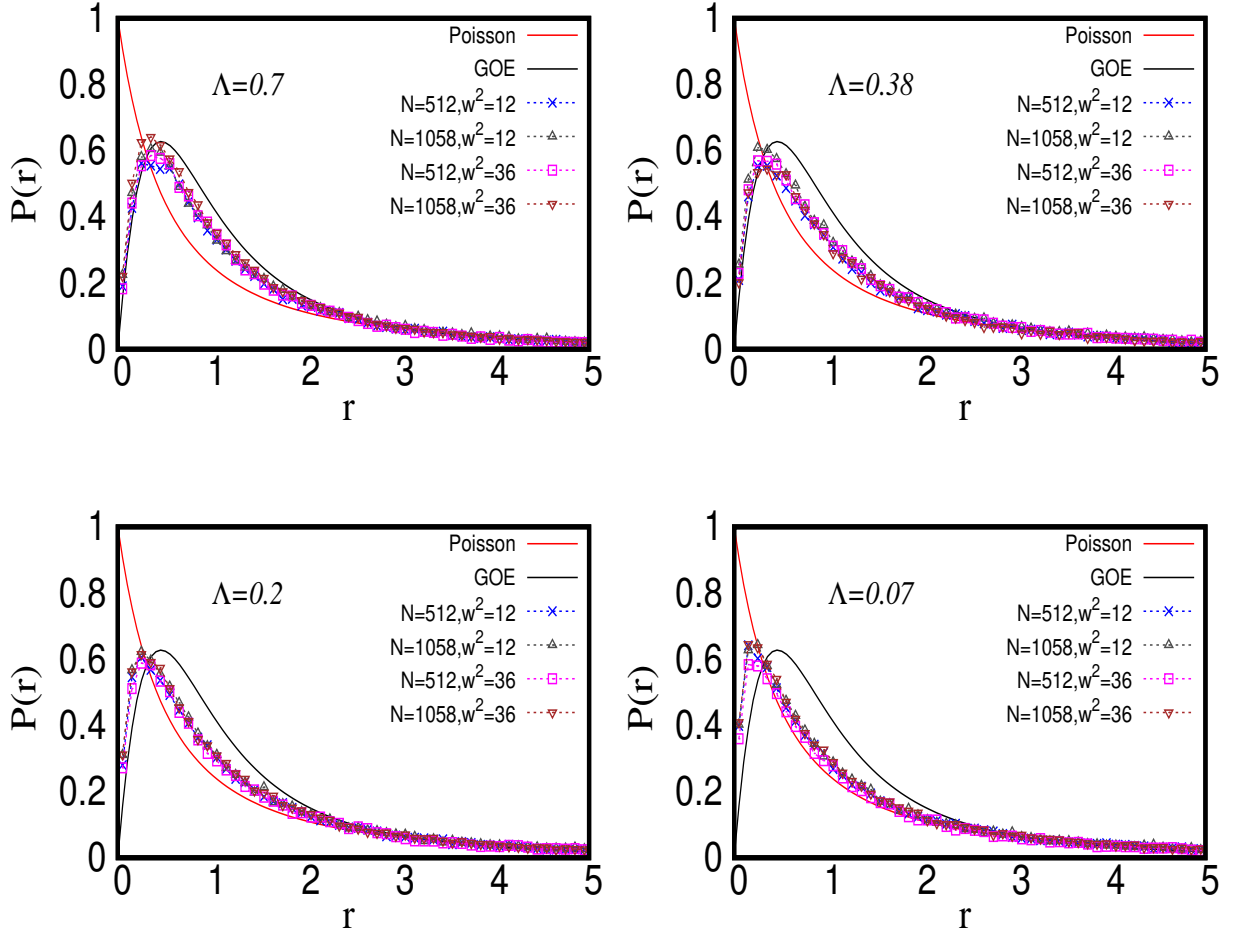}
	\caption{{\bf Intra-system analogy: nearest neighbour spacing ratio distribution for Ch-AE}: As $\Lambda_e$ for Ch-AE depends on many system parameters e.g. $ w, w_s, t$ and $N$, their different combinations  can result in same $\Lambda_e$ value. (see eq.(\ref{almu}) The figure here displays the analogies for four different $\Lambda_e$-values, latter spanning from Poisson ($\Lambda_e \to 0$ to Ch-GOE ($\Lambda_e \to \infty$) type spectral statistics.  Each part of the figure displays the $P(r)$ behavior for different Ch-AEs, corresponding to four different combinations of $e, w, N$ (keeping $w_s$ and $t$ fixed) which keeps their $\Lambda_e$ equal.  The theoretical limits of Poisson ($\Lambda_e=0$) and GOE ($\Lambda_e=\infty$) is also shown for comparison. The convergence of $P(r)$ for 
each case to the same curve and for all values of $\Lambda_e$ lends support to our theoretical prediction about the latter being the only parameter governing the spectral fluctuations.  The details of system and spectral parameters used here are given in table I.  The numerical results for $P(r)$ for each $\Lambda$ value are also fitted with eq.(\ref{pr}) with $\alpha(\beta_t)=\Lambda-1.69\;(1-\beta_t) +0.89\;(1-\beta_t)^5$ and the fitting parameters $\beta_t$ and $C_t$ given in Table I; for clarity, the fitted curves are not displayed here.}      
\label{fig3}
\end{figure}

\begin{figure}[h!]
	\centering
	\includegraphics[width=1.2\textwidth, height=1.0\textwidth]{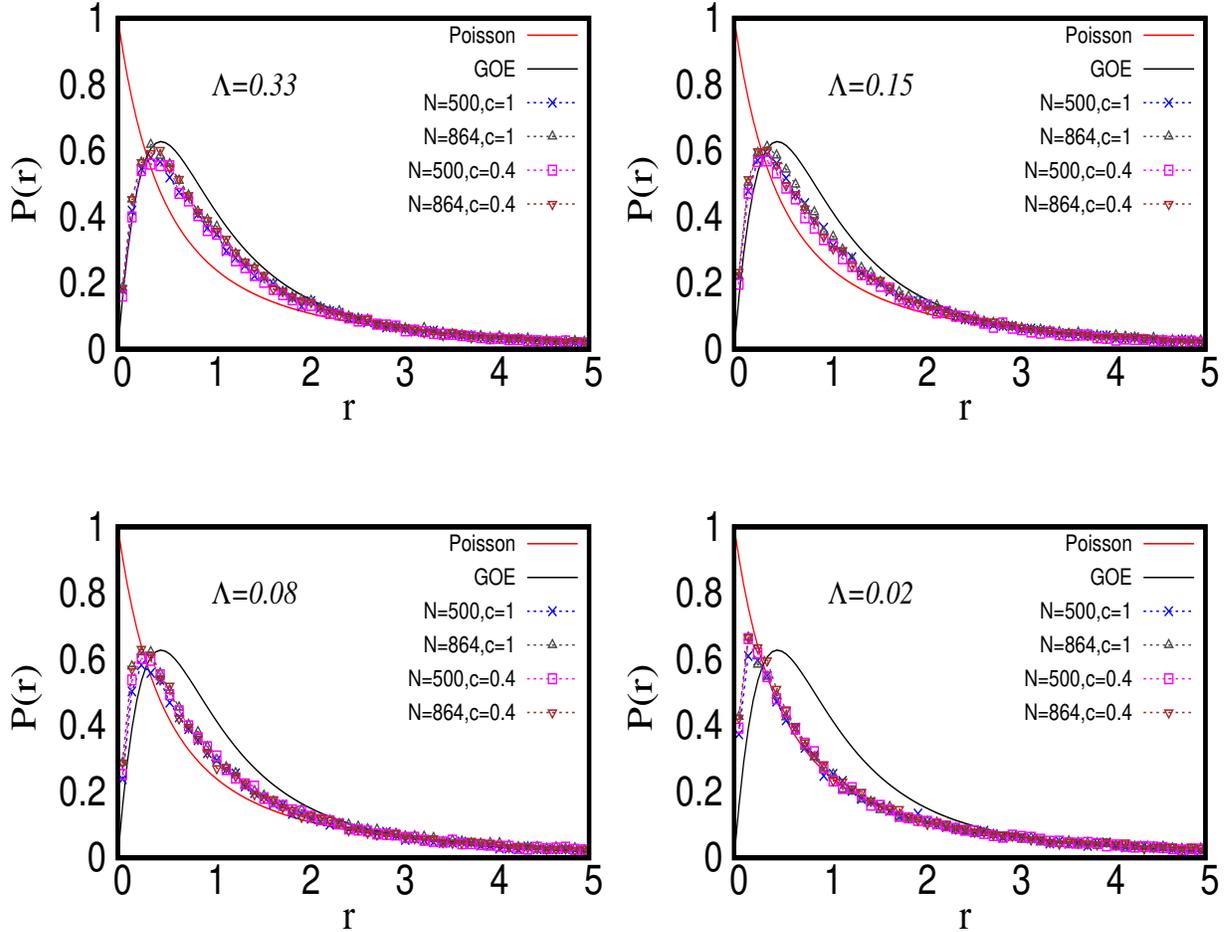}
	\caption{{\bf Intra-system analogy: nearest neighbour spacing ratio distribution for Ch-BE}: The details here are same as in figure 3 but now only two parameters namely $c, N$ are available to achieve same $\Lambda_e$-value for different Ch-BEs. The details of system and spectral parameters used here are given in table II.}
	\label{fig4}
\end{figure}

\begin{figure}[h!]
	\centering
	\includegraphics[width=1.2\textwidth, height=1.0\textwidth]{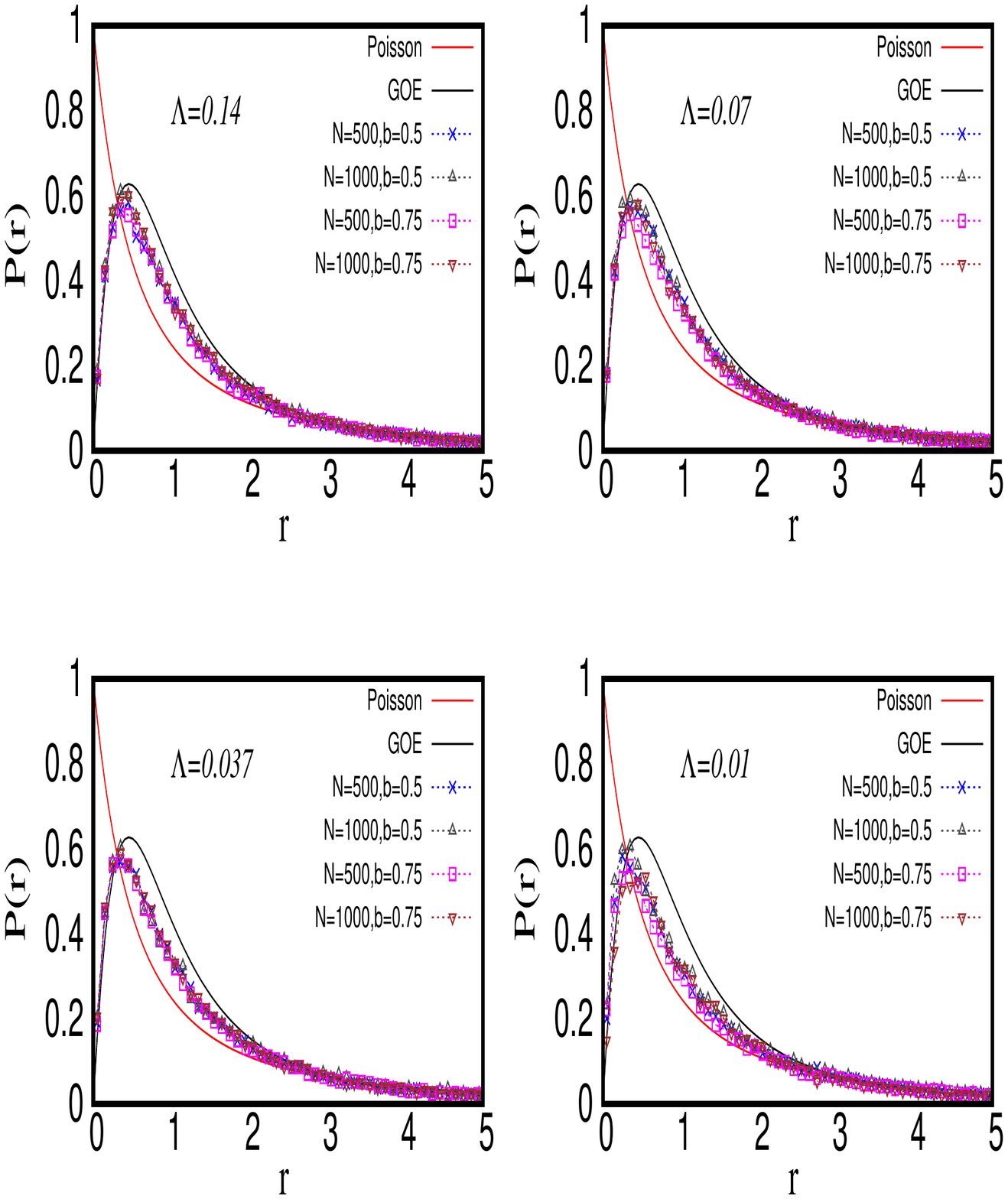}
	\caption{{\bf Intra-system analogy: nearest neighbour spacing ratio distribution for Ch-PE}: same as in figure 3 but here again only two parameters namely $b, N$ are available to achieve same $\Lambda_e$-value for different Ch-PEs. The details of system and spectral parameters used here are given in table III.}
	\label{fig5}
\end{figure}

\begin{figure}[h!]
	\centering
	\includegraphics[width=1.2\textwidth, height=1.0\textwidth]{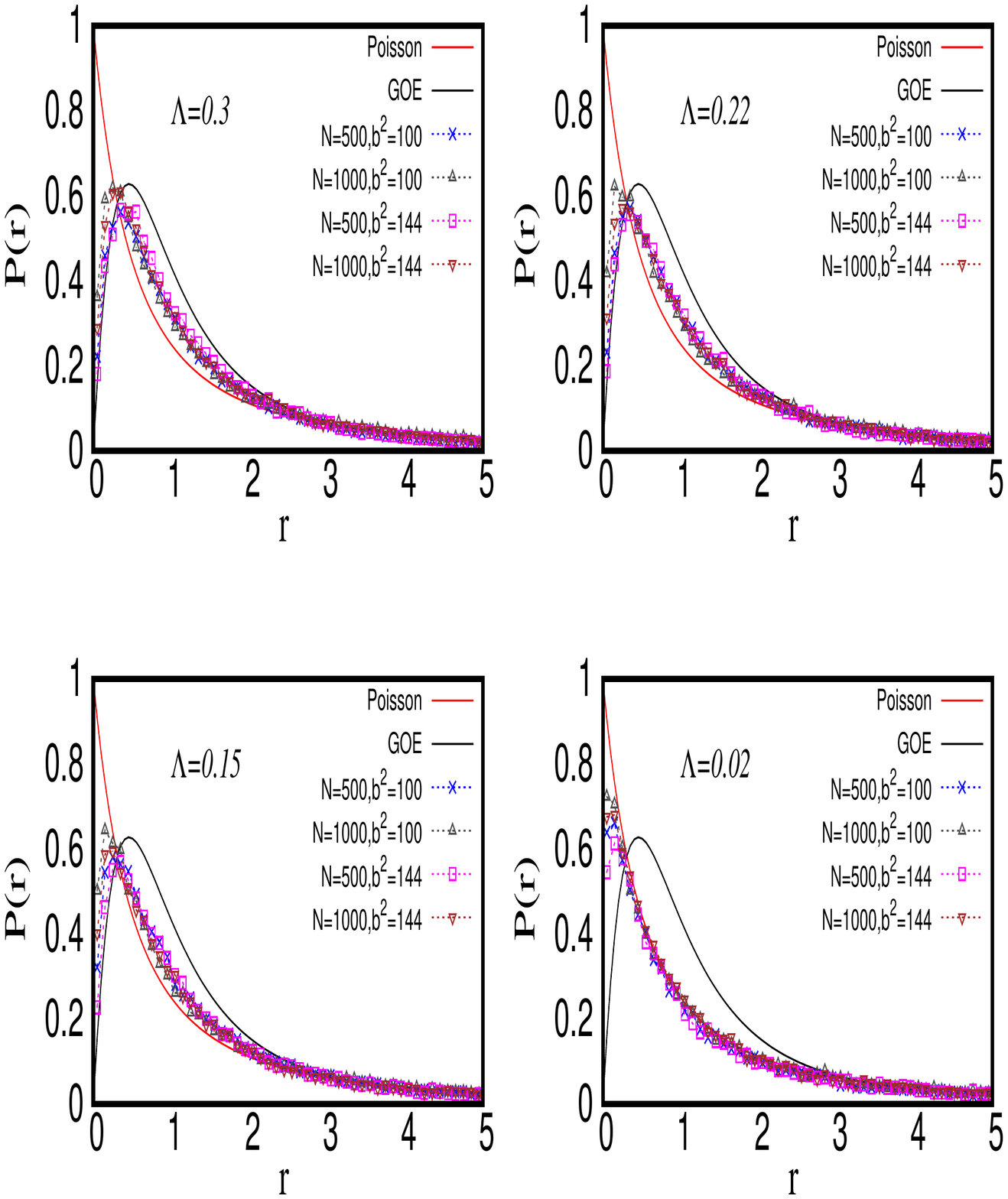}
	\caption{{\bf Intra-system analogy: nearest neighbour spacing ratio distribution for Ch-EE}: same as in figure 3 but here again only two parameters namely $b, N$ are available to achieve same $\Lambda_e$-value for different Ch-EEs. The details of system and spectral parameters used here are given in table IV.}
\label{fig6}
\end{figure}

\begin{figure}[h!]
	\centering
	\includegraphics[width=1.2\textwidth, height=1.0\textwidth]{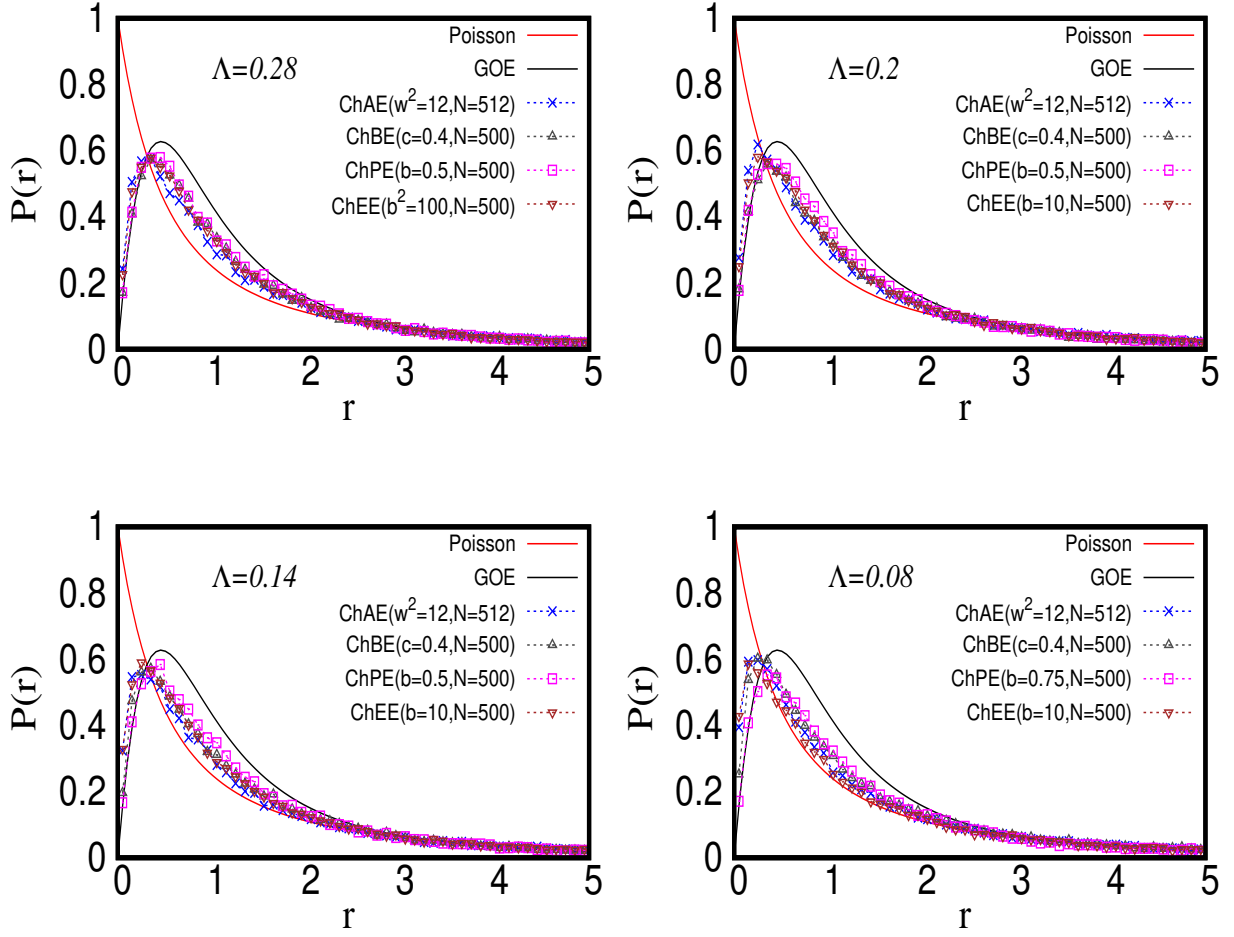}
	\caption{{\bf  Inter system analogy: nearest neighbour spacing ratio distribution for $\beta=1$}: The figure displays the comparison of  $P(r)$ for four different ensembles, namely, Ch-AE, Ch-BE, Ch-PE and Ch-EE.  Using  eq.(\ref{alma},\ref{almb},\ref{almp}, \ref{alme}), respectively, the system parameters in each case are chosen such that they lead to same $\Lambda_e$ (see eq.(\ref{almu1}). A good convergence of $P(r)$ for each case   
to the same curve for large $\Lambda$, and, an almost convergence for small $\Lambda$,  once again confirms the insensitivity of the spectral fluctuations to microscopic system details. The details of system and spectral parameters used here are given in table V.}
\label{fig7}
\end{figure}

\begin{figure}[h!]
\centering
\includegraphics[width=\textwidth, height=\textwidth]{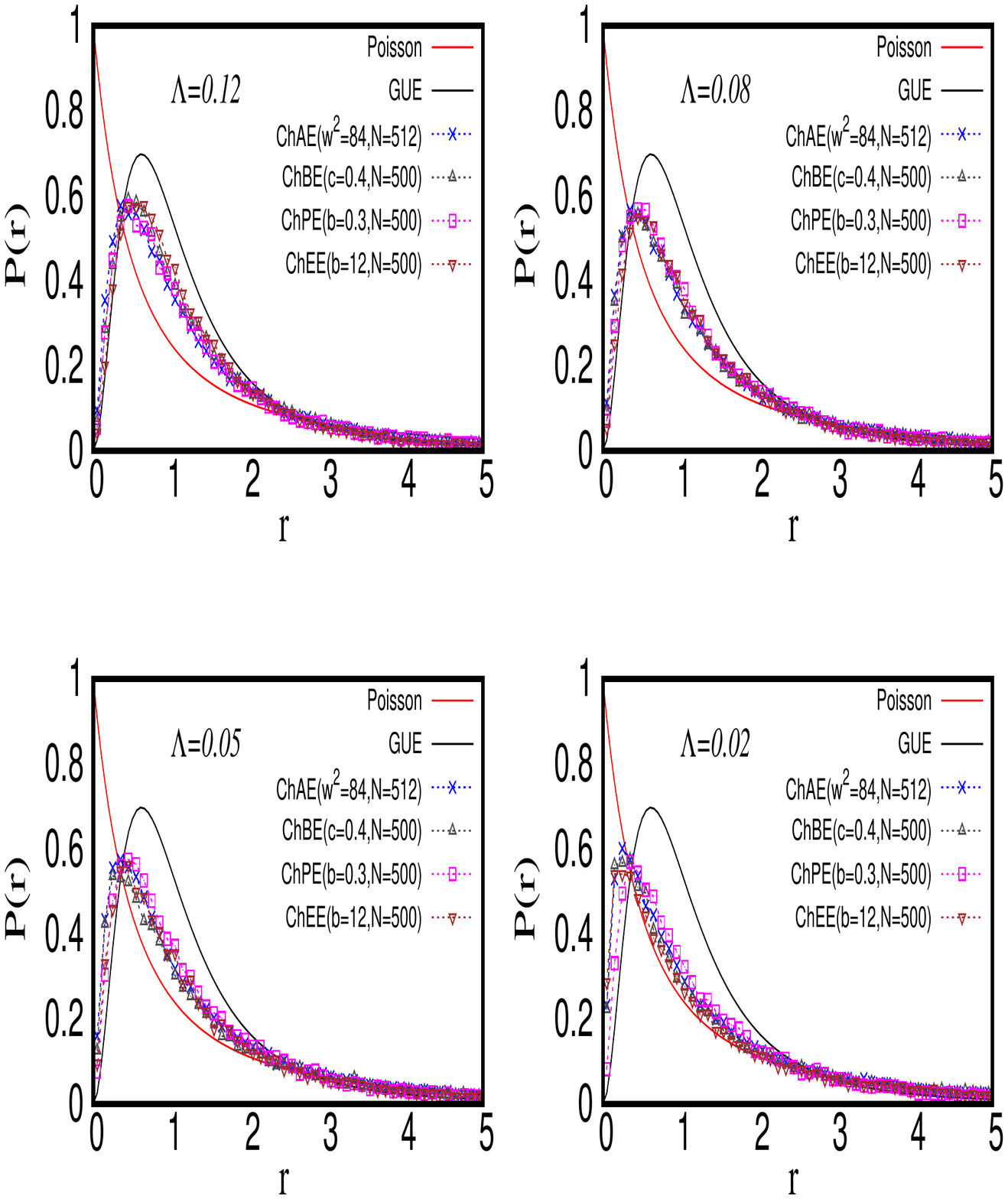}
\caption{{\bf Inter system analogy: nearest neighbour spacing ratio distribution for $\beta=2$}: As in figure 7, here again we compare the four ensembles but now $P(r)$ for each case is analyzed for $\beta=2$. The theoretical limits of Poisson ($\Lambda_e=0$) and GUE ($\Lambda_e=\infty$) are also shown for comparison. The other details are same as in figure 7 with details of system and spectral parameters used for this figure given in table VI.}
\label{fig8}
\end{figure}

\appendix

\section{Spectral response to change in system conditions}

The eigenvalue equation for the matrix $H$ can be given as $H U = { \lambda} U$ 
with ${ \lambda}$ as the diagonal matrix with eigenvalues $\lambda_i$ of $H$ as 
its matrix elements and $U$ as the eigenvector matrix (unitary for complex Hermitian case and orthogonal for real-symmetric case).

Assuming, the variation of system conditions leaves the chirality of $H$ unaffected, the eigenvalues $\lambda_{2N+k}$ for $k=1 \to \nu$ therefore remain zero throughout the dynamics. The dynamics of $\lambda_n$, with $n=1 \to N$, can then be given as
\begin{eqnarray}
\frac{\partial \lambda_n}{\partial H_{k,N+l;s}}=i^{s-1}[U_{kn}^* U_{N+l,n}+(-1)^{s+1} U_{N+l,n}^* U_{kn}]
\label{ch10}
\end{eqnarray}

The above in turn gives
\begin{eqnarray}
\sum_{k, l=1}^{N, N+\nu}\sum_{s=1}^{\beta} \frac{\partial \lambda_n}{\partial H_{k,N+l;s}}H_{k,N+l;s} &=& \lambda_n      \label{ch11} 
\end{eqnarray}

Further
\begin{eqnarray}
\sum_{k, l=1}^{N, N+\nu}\sum_{s=1}^{\beta} \frac{\partial \lambda_n}{\partial H_{k,N+l;s}}\frac{\partial \lambda_m}{\partial H_{k,N+l;s}}=\beta \; \delta_{mn} \quad {1 \le m,n \le N}\label{ch12}\\
\sum_{k, l=1}^{N,N+\nu} \sum_{s=1}^{\beta} \frac{\partial^2 \lambda_n}{\partial H^2_{k,N+l;s}}
= \frac{\beta(\nu+1/2)}{\lambda_n} + \sum_{m=1\neq n}^N \frac{2\; \beta \; \lambda_n}{\lambda_n^2-\lambda_m^2}
\label{ch14}
\end{eqnarray}

By replacing the subscript $n$ by ${N+n}$, the above equations can directly be used to derive  the response for the negative eigenvalues $\lambda_{n+N}$.  Note in this context that  chirality leads to the condition $U_{N+l,N+n} = -U_{N+l,n}$) and,  consequently, the replacement of  $\lambda_n \to - \lambda_{N+n}$ leaves  eq.(\ref{ch10},\ref{ch11},\ref{ch12},\ref{ch14})  invariant.

\section{Single Parametric Evolution of Eigenvalues}


	 The probability density $P_N(E,Y)$  of finding positive 
eigenvalues $\lambda_i $  between $e_i$ and $e_i+{\rm d}e_i$ at a given $Y$ can  be expressed in terms of the matrix elements distribution,
\begin{eqnarray}
P_N(E,Y)= \int
\prod_{i=1}^{N}\delta(e_i-\lambda_i) \; \delta(e_i+\lambda_{N+i}) \; { \rho(H,Y)} \; {\rm d}H 
\label{a1}
\end{eqnarray}
Here ${ E}$ refers to a diagonal matrix with elements $e_1,..,e_n$. 
 As the $Y$-dependence of $P$ in eq.(\ref{a1}) enters only through $\rho$, 
a derivative of $P$ with respect to $Y$  can be written as follows   

\begin{eqnarray}
{\partial P_N\over\partial Y}  
& = &  \int \prod_{i=1}^N \delta(e_i-\lambda_i) 
{\partial { \rho} \over \partial Y} {\rm d}H 
\label{a2}
\end{eqnarray}

Substitution of eq.(\ref{chi8}) in eq.(\ref{a2}) leads to 
\begin{eqnarray}
{\partial P_N\over\partial Y}  &=& I_1 +I_2 
\label{a3}
\end{eqnarray}
where 
\begin{eqnarray}
I_1 &=& \gamma \sum_{\mu} \int   \prod_{i=1}^N \delta(e_i-\lambda_i) \; \delta(e_i+\lambda_{N+i}) \;
 {\partial ({H_{\mu}  \; \rho})\over \partial H_{\mu}} \; {\rm d}H 
 \label{a4}\\
I_2 &=& \sum_{\mu} \int \prod_{i=1}^N \delta(e_i-\lambda_i) \; \delta(e_i+\lambda_{N+i}) \;
 {\partial^2 { \rho}\over \partial H_{\mu}^2}  \; {\rm d}H 
 \label{a5}
\end{eqnarray}
 with $H_\mu \equiv H_{k,N+l;s}$ with $1 \le k, l \le N$. 
 $I_1$ can further be simplified  by  integration by parts 
 \begin{eqnarray}
I_1 &=& -\gamma\sum_{\mu} \int {\partial \over \partial H_{\mu}} \left[
\prod_{i=1}^N \delta(e_i-\lambda_i) \; \delta(e_i+\lambda_{N+i}) \;\right]\; H_{\mu}\; { \rho} \;{\rm d}H \\ 
&=&2 \;  \gamma \sum_{n=1}^N {\partial \over \partial e_n} 
\int \prod_{i=1}^N \delta(e_i-\lambda_i) \; \delta(e_i+\lambda_{N+i}) \;
\left[\sum_{\mu} {\partial \lambda_n \over \partial H_{\mu}} 
 H_{\mu}\right] \; { \rho} \;{\rm d}H.  
 \label{a6}
\end{eqnarray}
Here the 2nd equality follows from the relations (i) ${\partial  \delta(e_n-\lambda_n)\over  \partial H_{\mu}} = {\partial \delta(e_n - \lambda_n)\over \partial e_n} {\partial \lambda_{n} \over \partial H_{\mu}}$, and, (ii)
${\partial  \delta(e_n+\lambda_{N+n})\over  \partial H_{\mu}} =- {\partial \delta(e_n+\lambda_{N+n})\over \partial e_n} {\partial \lambda_{N+n} \over \partial H_{\mu}} = {\partial \delta(e_n+\lambda_{N+n})\over \partial e_n} {\partial \lambda_{n} \over \partial H_{\mu}}$. Now using eq.(\ref{ch11}) of the appendix A in eq.(\ref{a6}), we have 
\begin{eqnarray}
I_1 &=& 2 \gamma \sum_{n}{\partial \over \partial e_n}\left(e_n P_N\right)
\label{a7}
\end{eqnarray}
$I_2$ can similarly be reduced as follows
\begin{eqnarray}
I_2 &=&  \sum_{\mu} 
\int {\partial^2 \over\partial H_{\mu}^2}\left(\prod_{i}\delta(e_i-\lambda_i) \; \delta(e_i+\lambda_{N+i}) \;\right) {\rho} \; {\rm d}H  \\
&=& 2\sum_{m,n} {\partial^2 \over \partial e_n \partial e_m}
\int\prod_{i}\delta(e_i-\lambda_i)\; \delta(e_i+\lambda_{N+i}) \;\left[ \sum_{\mu} 
{\partial \lambda_m \over \partial H_{\mu}} 
{\partial \lambda_n \over \partial H_{\mu}} \right]\; { \rho} \; {\rm d}H + \nonumber \\
&-& 2 \sum_{m} {\partial \over \partial e_n}
\int\prod_{i}\delta(e_i-\lambda_i) \; \delta(e_i+\lambda_{N+i}) \;\left[ \sum_{\mu} 
 {\partial^2 \lambda_n \over \partial H_{\mu}^2}\right]
\; { \rho} \; {\rm d}H \nonumber \\
\label{a8}
\end{eqnarray}
Further using eqs.(\ref{ch12}, \ref{ch14}), $I_2$ can be expressed  
in terms of eigenvalue derivatives of $\rho$,  
\begin{eqnarray}
I_2&=& 2 \; \sum_{n=1}^N \frac{\partial}{\partial e_n}\left[\frac{\partial}{\partial e_n}-\frac{(\nu+1/2)\beta}{ e_n}- \sum_{m=1}^N \frac{2 \beta e_n}{e_n^2- e_m^2} \right]P_N
\label{a9}
\end{eqnarray}
A substitution of $I_1$ and $I_2$ in eq.(\ref{a3}) leads to eq.(\ref{ch19})
describing the single parametric evolution of the eigenvalues of the
ensemble $\rho(H)$.

\newpage

\begin{table}[h!]
\centering
\caption{\textbf{Ensemble  and spectral parameters used for $\Lambda_e$ in Figure 3 for Ch-AE:} Here the first column lists the approximate $\Lambda_e$ values, referred as $\Lambda$  in different parts of  figure 3. The columns 2,3,4 list the ensemble parameters i.e matrix size, ensemble size and squared disorder strength $w^2 $ used in eq.(\ref{vand}) and eq.(\ref{alma}) with fixed $w_s^2=12$ and $t=0$. The columns 5,6,7 list the spectral parameters, namely,  energy $e$, ensemble averaged level density $R_1(e)/N$ and  inverse participation ratio at $e$ used in eq(\ref{alma}) with resulting $\Lambda_e$ values given in column 8. Here $\gamma=1/4$ fixed for all cases. We also compare the numerical results for $P(r)$ for each $\Lambda$ value with eq.(\ref{pr}) with $\alpha(\beta_t)=\Lambda-1.69\;(1-\beta_t) +0.89\;(1-\beta_t)^5$;
The columns 9, 10 list the values of fitted parameters $\beta_t$ and $C_t$. } 
\begin{tabular}{|c|c|c|c|c|c|c|c|c|c|c| }
\hline
 \hspace{0.5cm}$\Lambda$\hspace{0.5cm} & \hspace{0.5cm}$N$\hspace{0.5cm} & ensemble size & \hspace{0.5cm}$w^2$\hspace{0.5cm} & \hspace{0.5cm}$-e$\hspace{0.5cm} & \hspace{0.5cm}$F(e)$\hspace{0.5cm} & \hspace{0.5cm}$\langle I_2(e) \rangle$ \hspace{0.5cm} & \hspace{0.5cm}$\Lambda_{cal}  \hspace{0.5cm}$& $\beta_t$ & $C_t$\\
 &&&& ($\sim$) & ($\sim$) & ($\sim$) & ($\sim$) & & \\
\hline
	& 512 & 5000 & 12 &  1.39 & 0.143 & 0.02 & 0.69 &&\\
0.7 & 1058 & 2500 & 12 &  1.11 & 0.147 & 0.014 & 0.72 & 0.88 & 3.6\\
	& 512 & 5000 & 36 &  0.04 & 0.2295 & 0.032 & 0.696 &&\\
	& 1058 & 2500 & 36 &  0.04 & 0.233 & 0.0224 & 0.71 &&\\
\hline
	 & 512 & 5000 & 12 &  1.84 & 0.135 & 0.025 & 0.39 &&\\
0.38 & 1058 & 2500 & 12 &  1.58 & 0.137 & 0.018 & 0.379 & 0.88 & 4.07\\
	 & 512 & 5000 & 36 &  0.425 & 0.14 & 0.026 & 0.39 &&\\
	 & 1058 & 2500 & 36 &  0.095 & 0.16 & 0.021 & 0.38 &&\\
\hline
	& 512 & 5000 & 12 &  2.2 & 0.124 & 0.0326 & 0.196  &&\\
0.2 & 1058 & 2500 & 12 &  1.91 & 0.13 & 0.023 & 0.209 & 0.8 & 4.216\\
	& 512 & 5000 & 36 &  1.45 & 0.1205 & 0.0312 & 0.202 &&\\
	& 1058 & 2500 & 36 &  0.71 & 0.131 & 0.0237 & 0.2 &&\\
\hline
	 & 512 & 5000 & 12 &  2.65 & 0.109 & 0.0486 & 0.068 &&\\
0.07 & 1058 & 2500 & 12 &  2.39 & 0.116 & 0.035 & 0.072 & 0.66 & 3.7\\
	 & 512 & 5000 & 36 &  2.42 & 0.103 & 0.045 & 0.071 &&\\
	 & 1058 & 2500 & 36 &  1.96 & 0.114 & 0.034 & 0.073 &&\\
\hline
\end{tabular}      
\label{ae}
\end{table} 

\newpage


\begin{table}[h!]
\centering
\caption{\textbf{Ensemble  and spectral parameters used for $\Lambda_e$ in Figure 4 for Ch-BE:} The details here are same as in table I, except now the ensemble and spectral parameters refer to 
eq.(\ref{ch23}) and eq(\ref{almb}). Here again $\gamma=1/4$ is fixed for all cases. Note: In this case, $\langle I_2 \rangle$ is not required for  $\Lambda_e$ calculation.\\
}
\begin{tabular}{|c|c|c|c|c|c|c|c|c|}
\hline
 \hspace{0.5cm}$\Lambda$\hspace{0.5cm} & \hspace{0.5cm}$N$\hspace{0.5cm} & ensemble size & \hspace{0.5cm}$c$\hspace{0.5cm} & \hspace{0.5cm}$-e$\hspace{0.5cm} & \hspace{0.5cm}$F(e)$\hspace{0.5cm} & \hspace{0.5cm}$\Lambda_{cal}\hspace{0.5cm}$& $\beta_t$ & $C_t$\\
 &&&& ($\sim$) & ($\sim$) & ($\sim$) && \\
\hline
	 & 500 & 5000 & 1 &  0.08 & 0.2856 & 0.326 &&\\
0.33 & 864 & 2890 & 1 &  0.09 & 0.2846 & 0.324 &1 &4.7\\
	 & 500 & 5000 & 0.4 &  1.35 & 0.179 & 0.32  &&\\
	 & 864 & 2890 & 0.4 &  1.345 & 0.181 & 0.327 &&\\
\hline
	 & 500 & 5000 & 1 &  1.19 & 0.195 & 0.152 &&\\
0.15 & 864 & 2890 & 1 &  1.22 & 0.195 & 0.152 & 0.93 & 4.82\\
	 & 500 & 5000 & 0.4 &  1.87 & 0.122 & 0.148 &&\\
	 & 864 & 2890 & 0.4 &  1.82 & 0.122 & 0.148 &&\\
\hline
	 & 500 & 5000 & 1 &  1.65 &  0.143 & 0.081 &&\\
0.08 & 864 & 2890 & 1 &  1.66 & 0.142 & 0.08  & 0.83 & 4.52\\
	 & 500 & 5000 & 0.4 &  2.14 & 0.0897 & 0.08 &&\\
	 & 864 & 2890 & 0.4 &  2.14 & 0.09 & 0.081 && \\
\hline
	 & 500 & 5000 & 1 &  2.3 & 0.073 & 0.021 &&\\
0.02 & 864 & 2890 & 1 &  2.34 & 0.072 & 0.02 & 0.6 & 3.3\\
	 & 500 & 5000 & 0.4 &  2.75 & 0.045 & 0.02 &&\\
	 & 864 & 2890 & 0.4 &  2.72 & 0.045 & 0.02 &&\\
\hline
\end{tabular}         
\label{be}
\end{table}

\newpage

\begin{table}[h!]
\centering
\caption{\textbf{Ensemble and spectral Parameters  used  in Figure 5 for Ch-PE:} The details here are same as in table I, except now the ensemble and spectral  parameters refer to 
eq.(\ref{ch25}) and  eq(\ref{almp}).}
\begin{tabular}{|c|c|c|c|c|c|c|c| c|c|}
\hline
 \hspace{0.5cm}$\Lambda$\hspace{0.5cm} & \hspace{0.5cm}$N$\hspace{0.5cm} & ensemble size & \hspace{0.5cm}$b$\hspace{0.5cm} & \hspace{0.5cm}$-e$\hspace{0.5cm} & \hspace{0.5cm}$F(e)$\hspace{0.5cm} & \hspace{0.5cm}$\langle I_2(e) \rangle$ \hspace{0.5cm} & \hspace{0.5cm}$\Lambda_{cal} \hspace{0.5cm}$& $\beta_t$ & $C_t$ \\
 &&&& ($\sim$) & ($\sim$) & ($\sim$) & ($\sim$) &&\\
\hline
	 & 500 & 5000 & 0.5 &  1.53 & 0.1807 & 0.0184 & 0.1405 &&\\
0.14 & 1000 & 2500 & 0.5 &  1.55 & 0.177 & 0.0125 & 0.146 & 1 & 5.32 \\
	 & 500 & 5000 & 0.75 &  2.29 & 0.123 & 0.0179 & 0.138 &&\\
	 & 1000 & 2500 & 0.75 &  2.3 & 0.118 & 0.01195 & 0.143 &&\\
\hline
	 & 500 & 5000 & 0.5 &  1.69 & 0.164 & 0.0228 & 0.075 &&\\
0.07 & 1000 & 2500 & 0.5 &  1.72 & 0.162 & 0.0163 & 0.072 & 1 & 5.434 \\
	 & 500 & 5000 & 0.75 &  2.42 & 0.108 & 0.022 & 0.0706 &&\\
	 & 1000 & 2500 & 0.75 &  2.46 & 0.10857 & 0.0159 & 0.068 &&\\
\hline
	  & 500 & 5000 & 0.5 &  1.833 & 0.148 & 0.0287 & 0.038 &&\\
0.037 & 1000 & 2500 & 0.5 &  1.85 & 0.147 & 0.02054 & 0.037 & 1 & 5.43 \\
	  & 500 & 5000 & 0.75 &  2.55 & 0.098 & 0.0276 & 0.036 && \\
	  & 1000 & 2500 & 0.75 &  2.54 & 0.0968 & 0.0191 & 0.037 &&\\
\hline
	 & 500 & 5000 & 0.5 &  2.04 & 0.1169 & 0.0433 & 0.0106 && \\
0.01 & 1000 & 2500 & 0.5 &  2.04 & 0.1217 & 0.0319 & 0.0106 & 0.95 & 5.09\\
	 & 500 & 5000 & 0.75 &  2.75 & 0.0785 & 0.0429 & 0.0098 &&\\
	 & 1000 & 2500 & 0.75 &  2.7 & 0.08 & 0.0287 & 0.011 &&\\
\hline
\end{tabular}      
\label{pe}
\end{table} 

\newpage

\begin{table}[h!]
\centering
\caption{\textbf{Ensemble and spectral Parameters  used  in Figure 6 for Ch-EE:} The details here are same as in table II, except now the ensemble and spectral parameters refer to 
eq.(\ref{ch27})  and eq(\ref{alme}). Note here again, $\langle I_2 \rangle$ is not needed for $\Lambda_e$-calculation.}
   \begin{tabular}{|c|c|c|c|c|c|c|c|c| }
\hline
 \hspace{0.5cm}$\Lambda$\hspace{0.5cm} & \hspace{0.5cm}$N$\hspace{0.5cm} & ensemble size & \hspace{0.5cm}$b^2$\hspace{0.5cm} & \hspace{0.5cm}$-e$\hspace{0.5cm} & \hspace{0.5cm}$F(e)$\hspace{0.5cm} & \hspace{0.5cm}$\Lambda_{cal}\hspace{0.5cm}$ & $\beta_t$ & $C_t$\\
 &&&& ($\sim$) & ($\sim$) & ($\sim$) && \\
\hline
	& 500 & 5000 & 100 &  0.116 & 0.085 & 0.297 && \\
0.3 & 1000 & 2500 & 100 &  0.11 & 0.084 & 0.29 & 0.856 & 4.17\\
	& 500 & 5000 & 144 &  0.12 & 0.0774 & 0.298 && \\
	& 1000 & 2500 & 144 &  0.127 & 0.0765 & 0.29 &&\\
\hline
	 & 500 & 5000 & 100 &  2.14 & 0.0734 & 0.221 &&\\
0.22 & 1000 & 2500 & 100 &  1.995 & 0.0734 & 0.222 & 0.836 & 4.16\\
	 & 500 & 5000 & 144 &  2.07 & 0.0674 & 0.225 &&\\
	 & 1000 & 2500 & 144 &  2.36 & 0.0667 & 0.222 &&\\
\hline
	 & 500 & 5000 & 100 &  4.86 & 0.0604 & 0.15  &&\\
0.15 & 1000 & 2500 & 100 &  4.93 & 0.061 & 0.153 & 0.733 & 3.87\\
	 & 500 & 5000 & 144 &  5.29 & 0.0555 & 0.153  && \\
	 & 1000 & 2500 & 144 &  5.41 & 0.0549 & 0.1508 && \\
\hline
	 & 500 & 5000 & 100 &  7.93 & 0.0216 & 0.019 &&\\
0.02 & 1000 & 2500 & 100 &  7.98 & 0.024 & 0.023 & 0.335 & 2.13\\
	 & 500 & 5000 & 144 &  8.75 & 0.01985 & 0.0196 && \\
	 & 1000 & 2500 & 144 &  8.66 & 0.0206 & 0.21 && \\
\hline
\end{tabular}      
\label{ee}
\end{table} 

\newpage
\begin{table}[h!]
\centering
\caption{\textbf{Ensemble and spectral Parameters  used  in Figure 7 for Inter-system analogy for $\beta=1$ case:} The details here are same as in table I, except now the parameter given in column 4 refers to the ensemble mentioned in column 2 (given by eq.(\ref{vand}), eq.(\ref{ch23}), eq.(\ref{ch25}), eq.(\ref{ch27})). Similarly  the spectral parameters given in columns 5,6,7 are used for $\Lambda_e$ calculation of the systems in column 2, (with their $\Lambda_e$ given by eq.(\ref{alma}),  eq.(\ref{almb}),  eq.(\ref{almp}),  eq.(\ref{alme})). Here the ensemble size  is kept fixed (with 5000 matrices) for all cases.}
\begin{tabular}{|c|c|c|c|c|c|c|c|c|c| }
\hline
 \hspace{0.5cm}$\Lambda$\hspace{0.5cm} & \hspace{0.5cm}System\hspace{0.5cm} & \hspace{0.5cm}$N$\hspace{0.5cm} &  Disorder parameter & \hspace{0.5cm}$-e$\hspace{0.5cm} & \hspace{0.5cm}$F(e)$\hspace{0.5cm} & \hspace{0.5cm}$\langle I_2(e) \rangle$ \hspace{0.5cm} & \hspace{0.5cm}$\Lambda_{cal} \hspace{0.5cm}$ & $\beta_t$ & $C_t$\\
 
 &&& & ($\sim$) & ($\sim$) & ($\sim$) & ($\sim$) &&\\
 
\hline
	 & Ch-AE & 512 & $w^2=$12 & 2 & 0.128 & 0.0281 & 0.28 &&\\
0.28 & Ch-BE & 500 & $c=$0.4 & 1.41 & 0.167 && 0.278 & 0.94 & 4.53\\
	 & Ch-PE & 500 & $b=$0.5 & 1.27 & 0.202 & 0.0145 & 0.282 &&\\
	 & Ch-EE & 500 & $b^2=$100 & 0.116 & 0.084 && 0.0289 &&\\
\hline
	& Ch-AE & 512 & $w^2=$12 & 2.2 & 0.124 & 0.0326 & 0.196 && \\
0.2 & Ch-BE & 500 & $c=$0.4 & 1.65 & 0.142 && 0.2  & 0.92 & 4.62 \\
	& Ch-PE & 500 & $b=$0.5 & 1.41 & 0.191 & 0.0163 & 0.2 && \\
	& Ch-EE & 500 & $b^2=$100 & 2.78 & 0.0715 && 0.209 && \\
\hline
	 & Ch-AE & 512 & $w^2=$12 & 2.36 & 0.1206 & 0.037 & 0.143 && \\
0.14 & Ch-BE & 500 &$c=$ 0.4 & 1.87 & 0.12 && 0.144 & 0.87 & 4.46\\
	 & Ch-PE & 500 & $b=$0.5 & 1.53 & 0.1807 & 0.0184 & 0.1405 &&\\
	 & Ch-EE & 500 & $b^2=$100 & 5.09 & 0.0582 && 0.139 && \\
\hline
	 & Ch-AE & 512 & $w^2=$12 &  2.59 & 0.113 & 0.0457 & 0.082 &&\\
0.08 & Ch-BE & 500 & $c=$0.4 & 2.14 & 0.0897 && 0.08 & 0.78 & 4.18\\
	& Ch-PE & 500 & $b=$0.75 & 2.39 & 0.112 & 0.021 & 0.083 &&\\
	 & Ch-EE & 500 & $b^2=$100 & 6.6 & 0.045 && 0.083 && \\
\hline
\end{tabular}      
\label{ae2}
\end{table}

\newpage
\begin{table}[h!]
\centering
\caption{\textbf{Ensemble and spectral Parameters  used  in Figure 8 for Inter-system analogy for $\beta=2$ case:} The other details here are same as given in caption of table V.}
\begin{tabular}{|c|c|c|c|c|c|c|c|c|c| }
\hline
 \hspace{0.5cm}$\Lambda$\hspace{0.5cm} & \hspace{0.5cm}System\hspace{0.5cm} & \hspace{0.5cm}$N$\hspace{0.5cm} &  Disorder parameter  & \hspace{0.5cm}$-e$\hspace{0.5cm} & \hspace{0.5cm}$F(e)$\hspace{0.5cm} & \hspace{0.5cm}$\langle I_2(e) \rangle$ \hspace{0.5cm} & \hspace{0.5cm}$\Lambda_{cal} \hspace{0.5cm}$ & $\beta_t$ & $C_t$\\
 
 &&& & ($\sim$) & ($\sim$) & ($\sim$) & ($\sim$) &&\\
 
\hline
	 & Ch-AE & 512 & $w^2$ = 84 & 4.36 & 0.0693 & 0.017 & 0.12 &&\\
0.12 & Ch-BE & 500 & c = 0.4 & 2.7 & 0.111 && 0.123 & 1.89 & 22.445\\
	 & Ch-PE & 500 & b = 0.3 & 1.67 & 0.1938 & 0.01316 & 0.121 &&\\
	 & Ch-EE & 500 & $b^2$ = 144 & 1.04 & 0.04917 && 0.12 &&\\
\hline
	 & Ch-AE & 512 & $w^2$ = 84 & 1.33 & 0.0693 & 0.0204 & 0.083 &&\\
0.08 & Ch-BE & 500 & c = 0.4 & 2.91 & 0.09 && 0.081 & 1.745 & 21.18\\
	 & Ch-PE & 500 & b = 0.3 & 1.83 & 0.1841 & 0.01548 & 0.0795 &&\\
	 & Ch-EE & 500 & $b^2$ = 144 & 7.26 & 0.0404 && 0.081 &&\\
\hline
	 & Ch-AE & 512 & $w^2$ = 84 & 0.98 & 0.06585 & 0.025 & 0.05 &&\\
0.05 & Ch-BE & 500 & c = 0.4 & 3.08 & 0.072 && 0.051 & 1.6 & 19.145\\
	 & Ch-PE & 500 & b = 0.3 & 1.96 & 0.1744 & 0.0183 & 0.051 &&\\
	 & Ch-EE & 500 & $b^2$ = 144 & 9.76 & 0.0318 && 0.05 &&\\
\hline
	& Ch-AE & 512 & $w^2$ = 84 & 5.63 & 0.0606 & 0.0349 & 0.02 &&\\
0.02 & Ch-BE & 500 & c = 0.4 & 3.46 & 0.053 && 0.028 & 1.26 & 13.7\\
	 & Ch-PE & 500 & b = 0.3 & 2.14 & 0.155 & 0.0248 & 0.021 &&\\
	 & Ch-EE & 500 & $b^2$ = 144 & 11.8 & 0.0202 && 0.02 &&\\
\hline
\end{tabular}      
\label{ae3}
\end{table}

\end{document}